\documentclass{iopjournal}
\usepackage{graphicx}
\usepackage[dvipsnames]{xcolor}
\usepackage{svg}
\usepackage{float}
\usepackage{amsmath}
\usepackage{amssymb}
\usepackage[version=4]{mhchem}
\usepackage[style=phys]{biblatex}
\usepackage{xspace}
\usepackage{placeins}

\graphicspath{ {./Figures/} }
\newcommand{\epsr}{$\mathrm{\varepsilon_r}$\xspace}
\newcommand{\modealpha}{$\mathrm{\alpha}$\xspace}
\newcommand{\modebeta}{$\mathrm{\beta}$\xspace}
\newcommand{\modegamma}{$\mathrm{\gamma}$\xspace}
\newcommand{\degC}{$^{\circ}\text{C}$\xspace}
\newcommand{\us}{$\mu$s\xspace}
\newcommand{\um}{$\mu$m\xspace}
\newcommand{\fps}{fps\xspace}

\newcommand{\TE}[1]{\ensuremath{\text{TE}_{#1}}\xspace}
\newcommand{\TM}[1]{\ensuremath{\text{TM}_{#1}}\xspace}
\newcommand{\HE}[2]{\ensuremath{\text{HE}_{#1}^{#2}}\xspace}

\begin{document}

\articletype{Paper} 

\title{Microwave Resonant Discharges for Spatiotemporally Selective Plasma Breakdown Near Surfaces}

\author{Arnav Mohapatra$^{1\dagger}$\orcid{0009-0003-2513-6924}, Joshua K. Goodrich$^{1\dagger}$ \orcid{0009-0009-8520-6027}, Usman Humayun$^{1}$, and Thomas C. Underwood$^{1, 2, *}$\orcid{0000-0001-5720-2568}}

\affil{$^1$ Department of Aerospace Engineering and Engineering Mechanics, The University of Texas at Austin, Austin, TX 78712, United States}\\
\affil{$^2$ Texas Materials Institute, The University of Texas at Austin, Austin, TX 78712, United States}\\
\affil{$^*$ Author to whom any correspondence should be addressed.}\\
\affil{$^\dagger$ These authors contributed equally to this work.}\\

\email{thomas.underwood@utexas.edu}


\begin{abstract}
    Generating non-equilibrium plasmas close to surfaces remains a significant challenge for conventional plasma sources. Existing plasma generation schemes create volumetric discharges with limited spatial selectivity that lead to inefficient energy deposition and poor coupling between reactive intermediates and nearby surfaces. This work establishes tailored resonant field enhancement as a mechanism for prescribing where plasma forms near dielectric surfaces through microwave excitation. In this approach, the geometry, refractive index, and packing configuration of dielectric materials define resonant field structures that interfere constructively and amplify electric fields locally. Plasma forms only within these resonant volumes where the amplified fields exceed the local breakdown threshold, while the surrounding gas remains below breakdown. Microwave pulse shaping then provides dynamic control over these modes and can be used to excite different families of resonances, determine where breakdown occurs, and reconfigure what locations microplasmas occupy from one pulse to the next. We validate this framework through theory, electromagnetic simulations, and experiments using a pair of high-permittivity dielectric resonators. These studies identify multiple resonant mode families, demonstrate dynamic repositioning of microplasmas between prescribed breakdown sites, quantify the ignition characteristics of each mode, and confirm that multiple resonant microplasmas remain confined to localized field-enhancement regions during a microwave pulse. Together, these results establish a framework for designing resonant dielectric materials that localize, reconfigure, and control atmospheric-pressure plasmas near surfaces.
\end{abstract}

\section{Introduction} \label{intro}
    Non-equilibrium plasmas provide a method to drive chemical and surface reactivity without needing to bulk heat and pressurize gaseous reactants. In non-equilibrium plasmas, high-energy electrons collide inelastically with neutrals in a working gas, producing ions, radicals, and vibrationally excited species, among other reactive intermediates. These species enable plasma-assisted chemistry, materials processing, and surface modification by lowering activation barriers and allowing gas-phase and surface reactions to proceed at mild thermodynamic conditions \cite{Bruggeman2017FoundationsPlasmas,Fridman2012PlasmaChemistry, Subhankar2026DecouplingConversion, Subhankar2026Plasma-EnabledConversion, Park2003SynthesisCVD, Mehta2019CatalysisReview, Mariotti2010MicroplasmasSynthesis, Subhankar2025Plasma-EnabledCoking,Feng2025MitigatingCo-reactants,Xiong2024In-flightPlasma}. However, the synergy between excited reactants and surfaces is limited by the collisional relaxation of species. This converts the internal energy of excited states into bulk gas heating before species can be transported to a catalytic surface. For example, in atmospheric pressure air, experiments have shown \ce{N_2} vibrational to translational relaxation occurs within $\sim 7$~\us, which enables diffusive transport of $\sim 10$~\um before energy is quenched into translational heating \cite{Millikan1963SystematicsRelaxation, Lo2014SpaceO2, Popov2022RelaxationStabilization}. Existing plasma generation schemes struggle to generate plasmas sufficiently close to surfaces to overcome these relaxation times at atmospheric pressure. Instead, they create volumetric plasmas with limited spatial selectivity or control over where breakdown occurs. This leads to inefficient energy deposition and poor coupling between reactive intermediates and surfaces.
    
\begin{figure}[!b]
        \centering
        \includegraphics[width=1\textwidth]{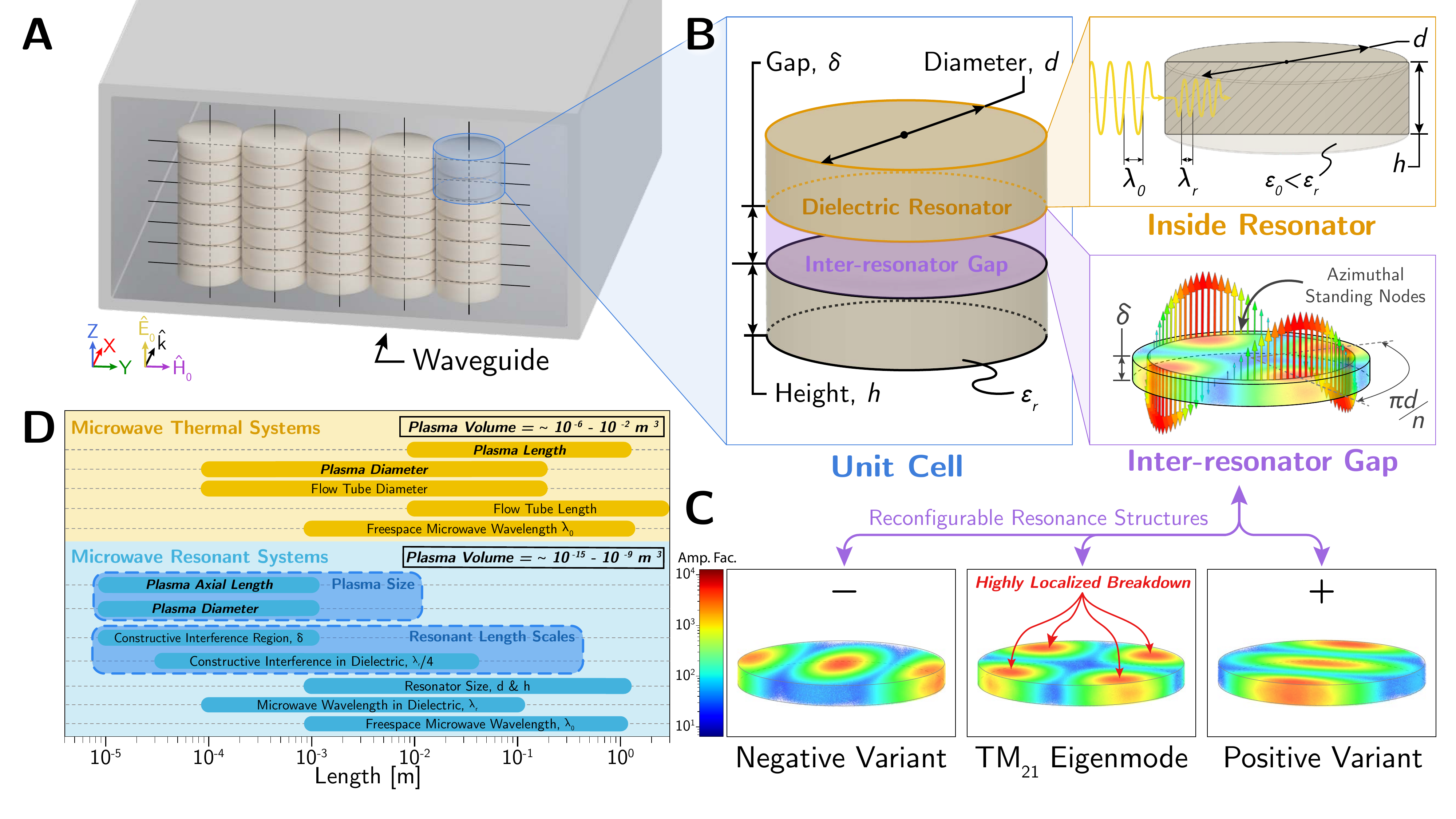}
        \vspace{-25pt}
        \caption{\textbf{A)} Visualization of a unit cell within a dielectric matrix as a simplified representation of coupled resonant elements in a microwave field. \textbf{B)} Geometric and material parameters, including diameter $d$, height $h$, gap $\delta$, and dielectric constant \epsr, that tune the resonant spectrum and field localization sites. \textbf{C)} The \TM{21} family and neighboring hybrid responses, showing how different resonant field structures can localize the electric field in different regions. \textbf{D)} Comparison of accessible plasma volumes relative to microwave thermal systems, highlighting the use of resonant field localization to confine breakdown to smaller prescribed regions.}
        \label{IntroFig1}
    \end{figure}

    Electromagnetic coupling is used in a variety of plasma sources to heat gases and generate volumetric plasmas without electrodes. These sources deposit energy at radio or microwave frequencies into partially ionized gas streams and can sustain large, high-density discharges across a wide range of pressures and gas compositions. Among these, microwave thermal (MWT) sources have become prevalent in atmospheric pressure gas stream processing for plasma-assisted chemical conversion \cite{Moisan1992MicrowavePlasmas, delaFuente2017MicrowaveProcesses, Trelles2020NonequilibriumFlows}. For MWT sources, the plasma location and structure are governed by applicator geometry, gas flow, thermal boundary conditions, and excitation frequency (Fig. \ref{IntroFig1}~D). However, sustaining a plasma at atmospheric pressure air requires an electric field of at least $\ge$3 MV/m \cite{Sun2014TheBreakdown}. These conditions result in excessive heating that can push gas temperatures in excess of $\sim$1200~K \cite{Nowakowska2013ModellingMPS}. Such high gas temperatures can sinter materials rapidly, deactivate catalysts, or otherwise damage surfaces that are enveloped within the discharge. These shortcomings make using MWT sources challenging for plasma generation near surfaces or within multiscale materials.

    Electrode driven systems, like dielectric barrier discharge devices (DBDs), address part of this limitation by generating non-equilibrium plasmas adjacent to dielectric surfaces at atmospheric pressures. In a DBD, an AC voltage applied across a dielectric-covered electrode gap generates plasma streamers where the reduced electric field (E/N) exceeds breakdown threshold \cite{Raizer1997GasPhysics}. Individual streamers are highly localized, with diameters of approximately 100 \um for millimeter-scale gaps \cite{Brandenburg2023BarrierUpdate}. Their growth is terminated by charge accumulation on dielectric surfaces, which reduces the electric field locally and quenches the discharge channel rapidly \cite{Kogelschatz2003Dielectric-BarrierApplications}. As a result, streamers are filamentary and transient, with each discharge forming at locations determined by the local surface charge, gas composition, electric field non-uniformity, and discharge history rather than at prescribed locations. This behavior is particularly limiting in structured dielectric media, where surfaces are treated as boundary conditions and rapidly accumulate charge. Consequently, streamers form preferentially at particle contact points and outer surfaces instead of uniformly penetrating the material interior \cite{Bogaerts2019BurningModeling, VanLaer2016FluidReactor, Zhang2016InfluencePores}. DBDs can therefore generate plasma near surfaces, but they cannot prescribe the location, duration, or structure of individual discharge events at specific surfaces of interest.

    High-Q resonators are an emerging scheme to generate plasmas by shaping where the electric field is amplified. These systems combine the advantages of electrode-less microwave excitation with the near-surface plasma production of DBDs \cite{ColonQuinones2018Laser-producedResonators, ColonQuinones2021TunableSpheroids}. Example microplasma devices have used split ring resonator arrays or plasma photonic schemes to generate microwave plasma breakdown in atmospheric pressure \cite{Hoskinson2016GasResonators}. Dielectric resonators build on this by leveraging the refractive and geometric properties of materials at microwave frequencies to amplify field strength over specific volumes \cite{Akram2026AnSource, Cohick2016AMetamaterials}. When a dielectric resonator is driven near one of its eigenmodes, the resonance forms a standing wave field pattern where constructive interference takes place (Fig. \ref{IntroFig1}~A-B) set by the index of the refraction of the material and its geometric length scales \cite{Kajfez1986DielectricResonators}. The high refractive index of the dielectric shortens the wavelength inside the resonator, so constructive interference can occur over dimensions that are fractions of the microwave wavelength. As this enhanced field travels through the dielectric walls and into the surrounding gas, electric fields can amplify by orders of magnitude to satisfy the dielectric-air boundary condition change \cite{Chelvam2017ModelingResonators}. When neighboring resonators are placed close together, these external fields overlap in the gap and can add up constructively (Fig. \ref{IntroFig1}~C) and concentrate the electric field in sub-wavelength scale regions \cite{Albella2013Low-lossDimers}. To date, dielectric resonances have primarily been explored as a means of enhancing electric fields and reducing the power required for plasma ignition. Their broader potential to localize plasmas near surfaces, dynamically reconfigure where breakdown occurs dynamically, and control multi-scale plasma-surface interactions has remained largely unexplored.

    In this paper, we establish tailored resonant field enhancement as a strategy for prescribing where plasma forms within dielectric materials and for reconfiguring those plasmas dynamically during operation. We show that geometry, packing configuration, and the refractive index of a dielectric assembly define multiple resonant field structures when excited by microwave frequencies. These resonant modes produce localized and spatially distinct electric-field amplification structures within specific voids and gaps (Fig. \ref{IntroFig1}~B). Plasma forms only within these resonant volumes where the amplified fields exceed the local breakdown threshold, while the surrounding material remains well below breakdown. As a result, microplasmas are generated over sub-wavelength length scales and are adjacent to dielectric surfaces rather than distributed through the entire volume. Microwave pulse shaping then provides dynamic control of these resonant microplasmas. Signal frequency selects which resonant field structure is excited to reposition the plasma between resonant volumes, high peak power determines when localized breakdown occurs, and average power controls dielectric and surface heating independently. Together, these capabilities provide a pathway to control how and where plasma forms near surfaces through microwave excitation without needing to modify the dielectric structure or its geometric arrangement.

    These capabilities were demonstrated using a pair of high-permittivity cylindrical resonators as a representative dielectric unit cell. Analytical theory, eigenmode analysis, and driven full-wave simulations identified families of resonant field structures, predicted their localized field amplification, and quantified how resonance frequencies evolve with dielectric geometry, packing configuration, and refractive index. A broadband high-power microwave source with independent control of frequency, pulse shape, and power was then used to excite these resonant modes experimentally. Frequency modulation during microwave pulsing demonstrated that distinct resonant modes could be excited sequentially to reposition microplasmas between prescribed subwavelength gaps, while peak-power modulation quantified the ignition and extinction thresholds of each mode. High-speed imaging confirmed that breakdown remained confined to resonant field-enhancement regions. These images also showed that microplasmas form at multiple resonance points simultaneously and remained confined to those volumes during microwave pulses. Together, these results establish a framework for designing resonant dielectric materials that localize, reconfigure, and control atmospheric-pressure plasmas near surfaces through microwave excitation.
   
    \FloatBarrier

\section{Theory} \label{theory}
    \subsection{Fundamental Eigenmodes, Driven Responses, and Spectral Self Similarity} \label{theory:eigenmodes}

        Electromagnetic resonances provide a way to store microwave energy in dielectric structures and shape where the electric field becomes concentrated. In a dielectric resonator, the resonant frequencies and field patterns are set by the geometry, permittivity, and surrounding boundary conditions. When the structure is driven near one of its eigenmodes, the resonance forms a standing wave field pattern associated with that mode. For a cylindrical dielectric resonator of diameter $d$, height $h$, and relative permittivity \epsr, the supported eigenmodes satisfy the Helmholtz equation with boundary conditions at the dielectric to air and air to waveguide interfaces. Treating the resonator as an isolated cylinder, the transverse magnetic modes of azimuthal order $m$ and radial order $n$ have resonant frequencies set by the transverse wavenumber inside the dielectric through the appropriate root $\chi_{mn}$ of the Bessel function $J_m$,
        
        \begin{equation}
            f_{r,\text{TM}} \approx \dfrac{c}{\pi \sqrt{\varepsilon_r}} \sqrt{\left(\dfrac{\chi_{mn}}{d} \right)^2 + \left(\dfrac{p \pi}{2h}\right)^2},
            \label{eqn:bessel_TM}
        \end{equation}
        
        \noindent where $c$ is the speed of light in vacuum and $p$ is the axial mode index \cite{Kajfez1986DielectricResonators}. At resonant frequencies, microwave energy is stored in the dielectric and organized into a spatial field pattern characteristic of the specified $m$, $n$, and $p$ mode. The dielectric to air boundary is only partially reflective, so part of the resonant field extends beyond the resonator surface into the surrounding gas. When two resonant dielectrics are brought into close proximity, these external fields overlap in the adjoining gas region and add constructively, further concentrating the electric field in the gap \cite{Albella2013Low-lossDimers}. Simulations of coupled microwave dielectric resonators have shown that this effect can increase the gap electric field to roughly 30 times the incident electric field \cite{Chelvam2017ModelingResonators}. This local enhancement allows the electric field to approach or exceed the microwave breakdown threshold of 3~MV/m at atmospheric pressure with much less incident power than would be required without resonant concentration \cite{Gongora-Nieto2003ImpactStrengths}. The coupled resonator geometry, therefore, provides a mechanism for generating plasma selectively within confined gas volumes.
        
        The resonator dimensions and permittivity were selected to place the targeted resonance family within the 2.8--3.5~GHz operating range of the waveguide and amplifier system. From Eq. \ref{eqn:bessel_TM}, a disk of 30~mm diameter, 8~mm height, and nominal \epsr=29 places the analytical \TM{21} resonance at 3.033~GHz. The first index of this mode denotes two full azimuthal field variations, producing the characteristic four lobed pattern, and the second denotes one radial variation from center to edge. Open dielectric resonators also support hybrid electromagnetic modes, commonly denoted HEM or HE modes, in which the fields do not reduce to a purely transverse magnetic or transverse electric solution \cite{Kajfez1986DielectricResonators,Mongia1994DielectricBandwidth}. The neighboring \HE{21}{-} and \HE{21}{+} modes are labeled as hybrid modes because they follow the same azimuthal and radial indexing convention as the central \TM{21} mode, while their field structures do not reduce to the pure \TM{21} pattern. The superscripts denote the lower and upper hybrid branches that lie on either side of the central \TM{21} resonance and track with it as part of the same resonance family.
    
\begin{figure}[!t]
            \centering
            \includegraphics[width=1\textwidth]{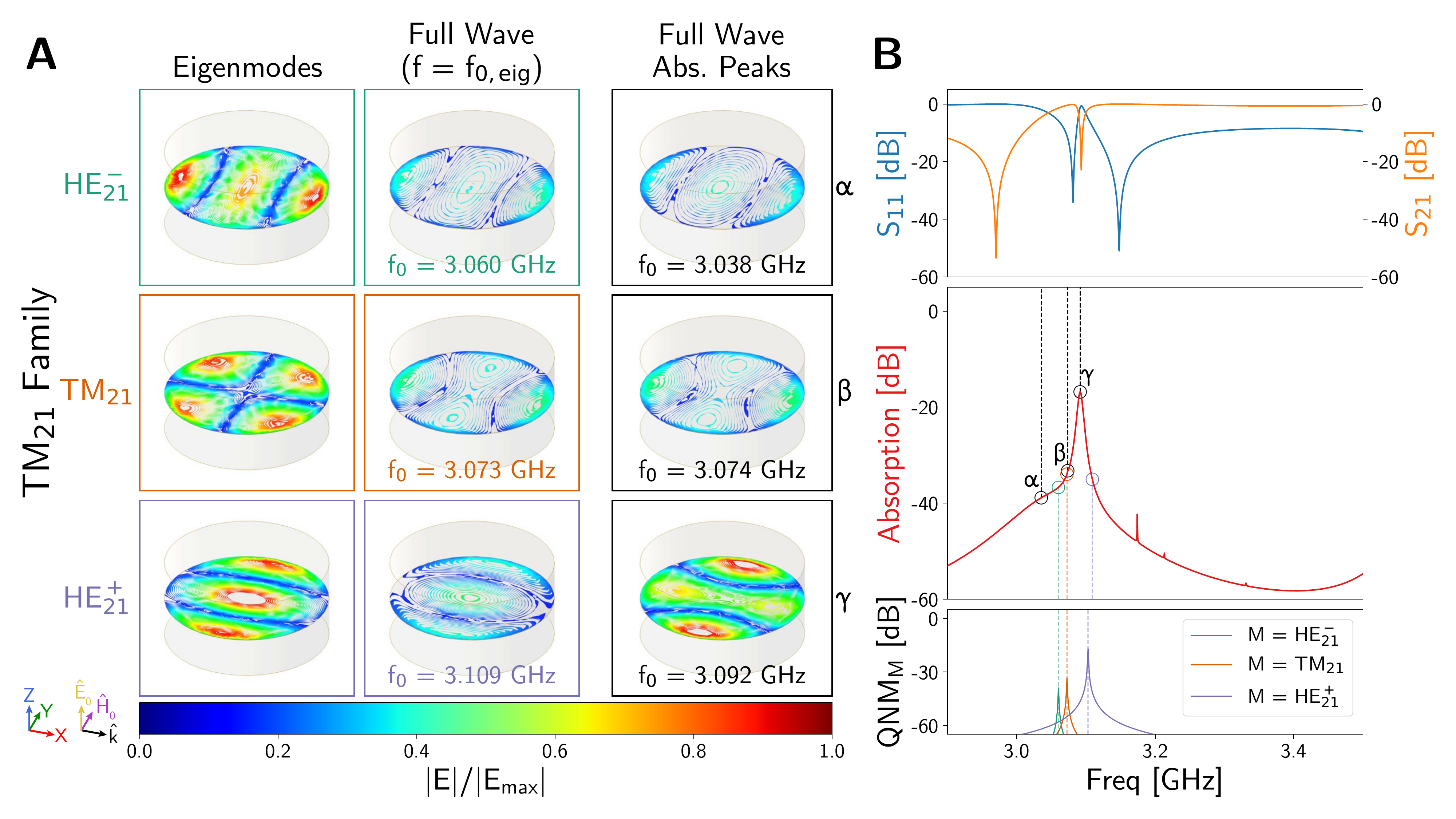}
            \vspace{-25pt}
            \caption{\textbf{A)} Electric field distributions of the \HE{21}{-}, \TM{21}, and \HE{21}{+} modes obtained from the eigenmode solver and driven full wave simulation. All field plots are normalized to the peak electric field of the highest absorption response. \textbf{B)} Corresponding $S_{11}$, $S_{21}$, and absorption spectra. The lower plot shows the individual quasi-normal mode contributions and their combined response, which produces the observed absorption features.}
            \label{SimFig1}
        \end{figure}

        To determine how these natural resonances appeared under experimental excitation, we modeled the coupled resonator system in ANSYS High Frequency Structure Simulator (HFSS) using two complementary approaches consisting of an eigenmode simulation and a driven full-wave(FW) simulation (Fig. \ref{SimFig1}). The eigenmode simulation placed the two resonators inside a closed perfect electric conductor cavity and solved the unforced problem for the natural resonances, yielding the complex resonant frequencies $f_{0,\text{eig}}$, quality factors $Q$, and modal field distributions. The driven full wave simulation placed the same geometry inside a WR-284 rectangular waveguide, excited the structure in the \TE{10} mode through matched ports, and swept frequency to obtain the scattering parameters $S_{11}$ and $S_{21}$ and the driven field at each frequency. The absorbed fraction of incident power follows from the scattering parameters as,
        
        \begin{equation}
            A=1-\left|S_{11}\right|^2-\left|S_{21}\right|^2.
            \label{eqn:Absorption}
        \end{equation}
        
        For the pair of disks separated by a 100~\um gap, the eigenmode solver returned three closely spaced resonances. These consisted of a lower hybrid \HE{21}{-} mode at 3.060~GHz, the central \TM{21} mode at 3.073~GHz, and an upper hybrid \HE{21}{+} mode at 3.109~GHz, where the superscripts denote relative spectral position below or above the central \TM{21} resonance. The \TM{21} eigenmode prediction lies about 40~MHz above the analytical value because the eigenmode model treats the setup as a loaded cavity, whereas the analytical solution solves the waveport's coupling to the loaded waveguide. 
        
        The driven full wave model produced responses at 3.038, 3.074, and 3.092~GHz that most closely resembled the \HE{21}{-}, \TM{21}, and \HE{21}{+} eigenmodes, respectively (Fig. \ref{SimFig1}). These responses are denoted \modealpha, \modebeta, and \modegamma to distinguish the driven field structures from isolated eigenmodes. The spectrally nearest eigenmode typically dominated each response, but the weaker superposition of neighboring modes slightly modified the field distribution and coupling strength.
        
        The three resonances lie close enough that their driven responses overlap into a broadened absorption feature rather than three isolated peaks (Fig. \ref{SimFig1}~B). Under driven excitation, the closely spaced eigenmodes superimpose and create absorption peaks that combine contributions from nearby modes, so the driven absorption peaks do not correspond one to one with isolated eigenmodes. Viewed along the waveguide axis, \modealpha shows a three region distribution concentrated at the front, middle, and rear of the gap, \modebeta shows the four lobed pattern of the $m=2$, $n=1$, $p=0$ \TM{21} solution, and \modegamma shows three regions at the left, middle, and right of the gap, skewed toward a four lobed structure by residual \TM{21} influence. With all fields normalized to the peak of the highest absorption response, \modegamma, the maximum fields of \modealpha and \modebeta are correspondingly lower, consistent with their smaller absorption maxima. For a fixed geometry and family, absorption magnitude therefore provides a useful proxy for the peak gap field and breakdown potential.
        
        
        \FloatBarrier
        
\begin{figure}[!t]
            \centering
            \includegraphics[width=1\textwidth]{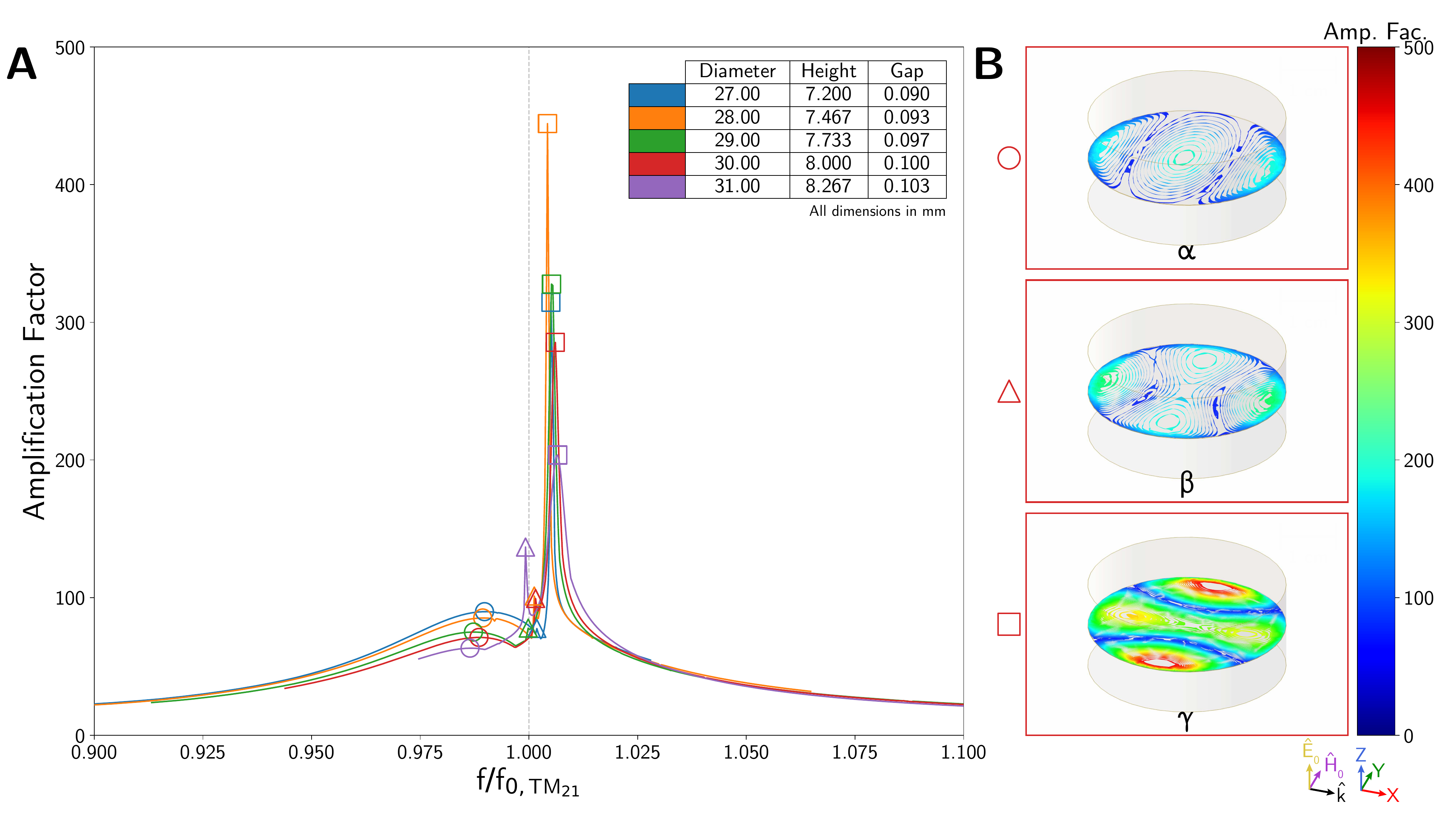}
            \vspace{-25pt}
            \caption{\textbf{A)} Amplification factor spectra for five geometrically scaled resonator configurations plotted against excitation frequency normalized by the analytically calculated \TM{21} resonance frequency. The normalized spectra demonstrate the approximate self-similar scaling of the coupled resonance family. \textbf{B)} The three primary amplification maxima identified for each geometric configuration, together with representative electric field distributions for the nominal geometry. The persistence of the modal field structures across the scaled configurations demonstrates that the resonance family remains identifiable as the characteristic length scale changes.}
            \label{SimFig3}
        \end{figure}

        To determine whether this resonance family remained consistent as the characteristic length scale changed, we compared the nominal parameter set to geometrically rescaled variants. This scaling test evaluated the Helmholtz behavior consistency implied by Eq. \ref{eqn:bessel_TM} and determined whether the relative ordering and field structures of the coupled resonance family persisted as the resonator size changed. Five geometries with disk diameters from 27 to 31~mm were simulated at fixed aspect ratio by scaling height and gap to maintain the $d:h:\delta$ ratio of $30 : 8 : 0.1$ (Fig. \ref{SimFig3}~A). Here, the amplification factor was defined as the maximum electric field in the gap normalized by the incident electric field amplitude. The driven amplification spectrum of each case was plotted against the normalized frequency $f_{\text{norm}} = f/f_{0,\text{TM}_{21}}$, where $f_{0,\text{TM}_{21}}$ is the analytical \TM{21} frequency defined by Eq. \ref{eqn:bessel_TM}. This normalization centered the spectra near the \TM{21} dominated \modebeta response (Fig.~\ref{SimFig3}~A).
        
        Electric field distributions extracted at each peak verified that the same field structures were tracked across the scaled configurations (Fig. \ref{SimFig3}~B). As expected, the \modebeta peaks collapsed to a narrow region near $f/f_{0,\text{TM}_{21}}\approx1$ for all five geometries, confirming that the analytical \TM{21} frequency tracked the center of the coupled family for all scales. The \modealpha and \modegamma responses remained on the left and right of \modebeta, respectively, preserving the \modealpha--\modebeta--\modegamma ordering at every size. This establishes that the resonance family, its ordering, and its field structures remain approximately self-similar when designing the resonance structure. Placing \modebeta near a desired frequency through Eq. \ref{eqn:bessel_TM} also placed \modealpha and \modegamma in predictable neighboring regions. This provides a pathway to estimate resonator dimensions from a single analytically tractable target frequency rather than searching broadband FW simulations for the entire family.
        
    
        \FloatBarrier
        
    \subsection{Impact of Gap on Resonances} \label{theory:gap_control}

        When manufacturing dielectric disks, parameters like diameter, height, and relative permittivity are locked in to a final value. The only remaining physical control parameter is gap, so understanding how this affects resonance is crucial to assembling and operating the system. A gap sweep was simulated to determine how the separation between dielectrics impacts the \modealpha, \modebeta, and \modegamma responses and the field strength available for plasma breakdown. As the disks approach contact, the effective electromagnetic discontinuity between them weakens, the fields merge, and the pair begins to respond like a single dielectric element with twice the individual disk height.

\begin{figure}[!t]
            \centering
            \includegraphics[width=1\textwidth]{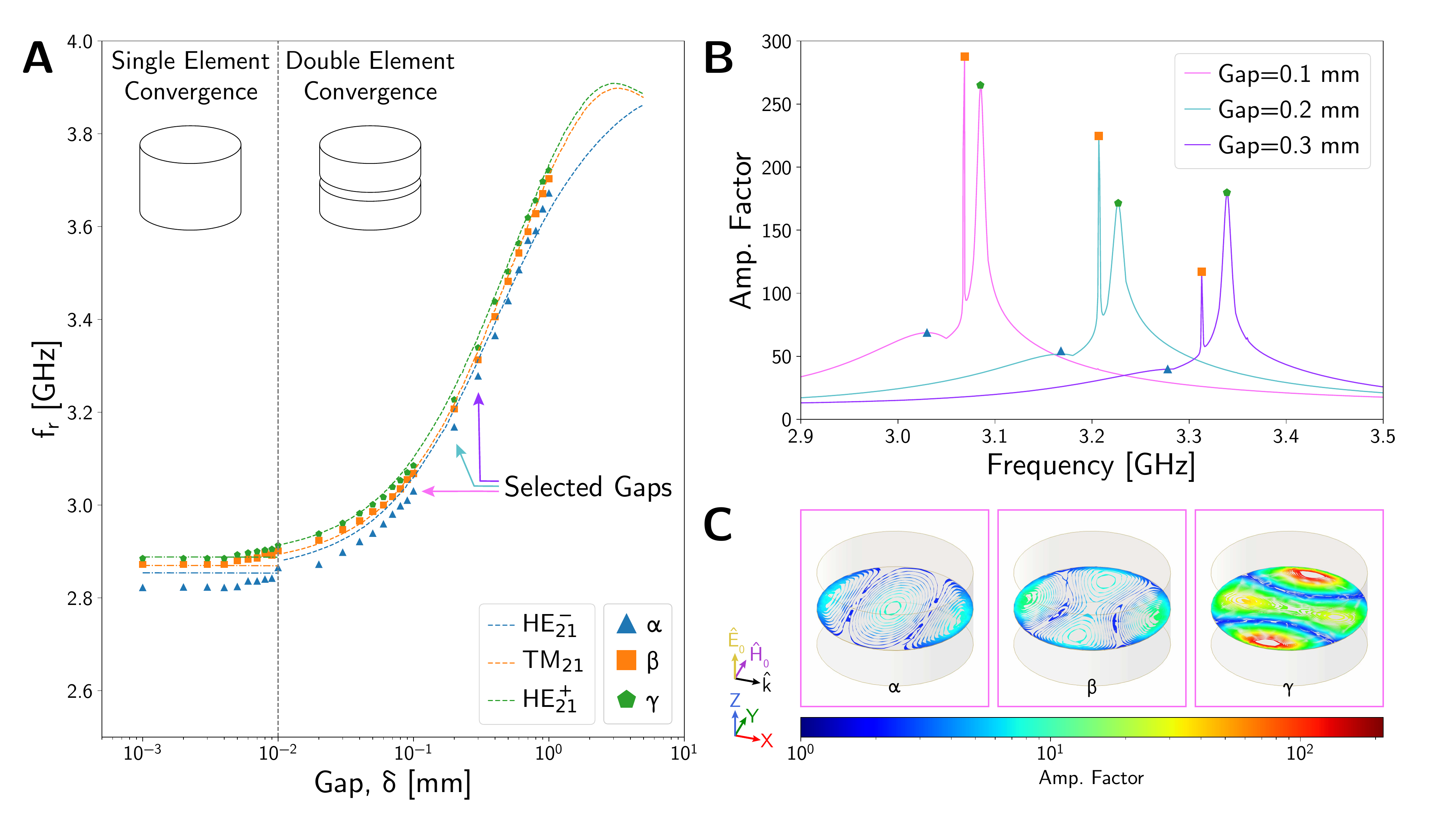}
            \vspace{-25pt}
            \caption{\textbf{A)} Resonant frequencies of the \HE{21}{-}-, \TM{21}-, and \HE{21}{+}-dominated responses, corresponding to \modealpha, \modebeta, and \modegamma, as functions of inter-resonator gap. Eigenmode analysis of a single element in the near-contact range from 1 to 10~\um gaps and of double-elements in the 10~\um to 4~mm gap range are shown in dashed lines. Amplification peaks of the driven full-wave model are shown as points corresponding to their mode structure. Agreement of full-wave and eigenmode simulations below 10~\um confirms the convergence into an effectively single-element solution at small gaps. \textbf{B)} Amplification factor spectra for gaps of 100, 200, and 300~\um, with the modal peaks identified for each configuration. \textbf{C)} Electric field distributions in the inter-resonator gap at the \modealpha, \modebeta, and \modegamma peaks for the 100~\um configuration.}
            \label{SimFig2}
        \end{figure}

        Simulations were completed to find resonant frequency $f_r$ at various gaps represented by mode-ordered peak amplification points and third-order polynomial fit dashed lines for FW and eigenmode simulations, respectively (Fig. \ref{SimFig2}~A). The near-contact limit was evaluated by comparing sweeps over gaps from 1 to 10~\um against double-height single-element eigenmode simulations with the same effective height. Simulations then swept the physical gap in discrete steps from 10~\um to 1~mm and extracted the \HE{21}{-}-, \TM{21}-, and \HE{21}{+}-dominated frequencies at each spacing. Polynomial fits for eigenmode simulations are included only as visual guides to gauge eigenmode-FW deviation. To quantify the plasma relevant field enhancement, the driven amplification factor was calculated at gaps of 100, 200, and 300~\um as the ratio of peak gap field to incident field (Fig. \ref{SimFig2}~B).

    
        The near-contact sweep shows that the double-element solution converges toward the single-element solution as the gap decreases from 10 to 1~\um. This confirms that, as the gap closes, the electromagnetic discontinuity between the disks effectively vanishes and the pair responds as if it were a single monolithic resonator with the combined height of each element. This provides a lower bound for gap tuning that ensures that there is sufficient separation for the two disks to remain distinct coupled resonators. Above 10~\um, the three resonant frequencies rise with increasing gap, moving the family away from the single-element limit \cite{Kajfez1986DielectricResonators}. The modes do not shift at identical rates, but they remain spectrally close and preserve the \HE{21}{-} $<$ \TM{21} $<$ \HE{21}{+} frequency ordering across the gap range. Electric field distributions extracted in the gap verify that the corresponding \modealpha, \modebeta, and \modegamma field structures remain identifiable across the parameter variation (Fig. \ref{SimFig2}~C).
    
        The amplification spectrum mirrors the absorption response identified previously, but more directly quantifies the peak electric field available for breakdown. At 100~\um, the resonances remain closely spaced, and the maximum amplification factor reaches approximately 290. This large amplification occurs as the near coincidence of the \TM{21}- and \HE{21}{+}-dominated responses allows their fields to overlap and shape the combined spectrum. Increasing the gap to 200 and 300~\um shifts all three peaks upward in frequency and lowers their amplification, as weaker coupling reduces the concentration of the amplified field in the inter-resonator region. The same \modealpha, \modebeta, and \modegamma field patterns persist across the full range, showing that the gap tunes the frequencies, amplitudes, and degree of modal overlap for the underlying resonance family. This identifies gap control as a control of the field amplification available to each resonant mode. For a fixed incident power, this will later determine which modes can ignite when which localized regions in the gap overcome the gas breakdown threshold.

        \FloatBarrier

    \subsection{Resonant Frequency and Q Factor Response to Resonator Parameter Variation} \label{theory:parameter_sweeps}

\begin{figure}[!t]
            \centering
            \includegraphics[width=1\textwidth]{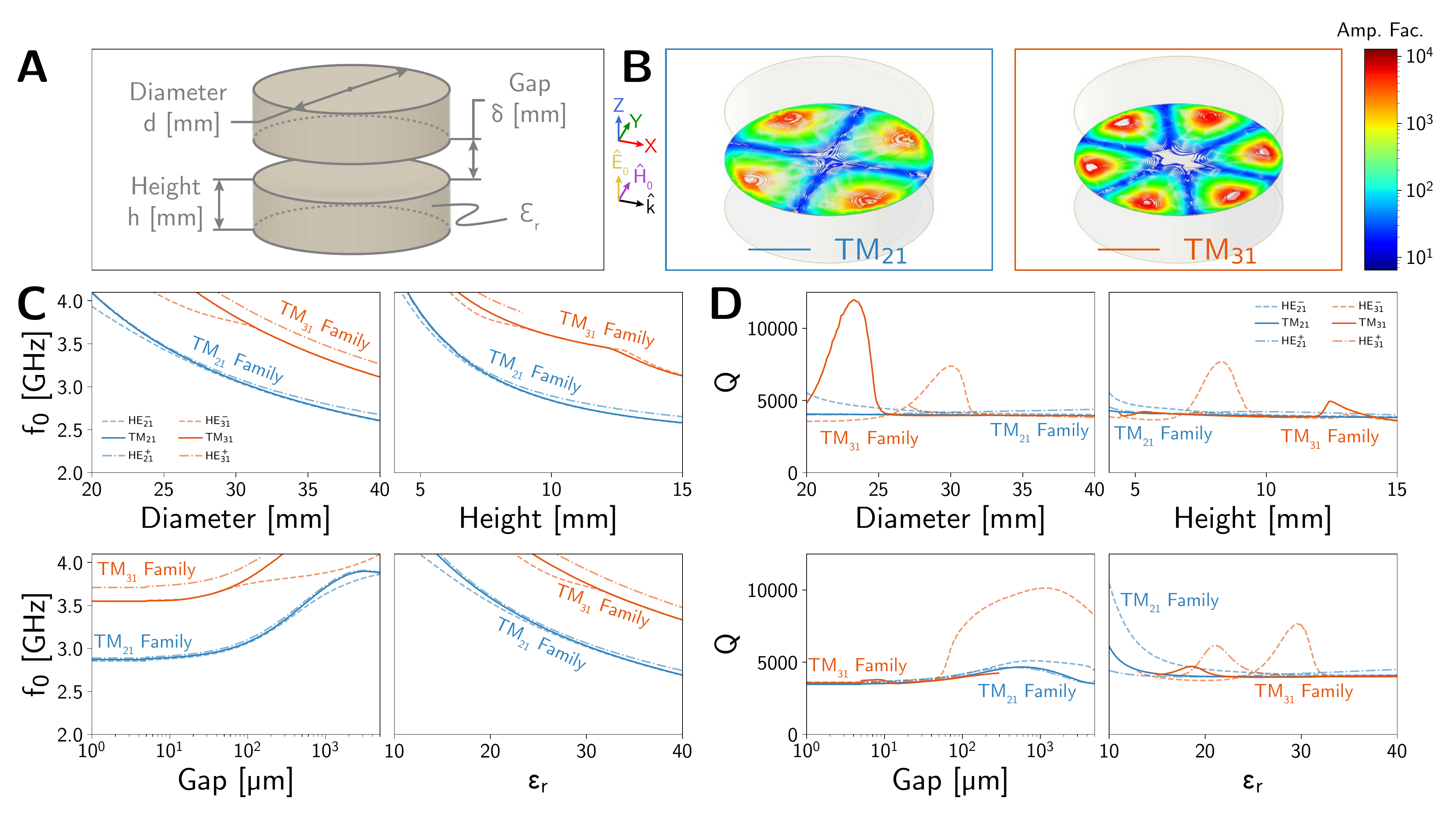}
            \vspace{-25pt}
            \caption{\textbf{A)} Coupled-resonator unit cell with the independently varied geometric and dielectric parameters identified. \textbf{B)} Representative electric-field distributions in the interaction region for the \TM{21} and \TM{31} eigenmodes. \textbf{C)} Resonant-frequency response to independent variations in resonator diameter, height, inter-resonator gap, and relative permittivity. \textbf{D)} Corresponding quality-factor response. Each sweep varies one parameter while holding the remaining parameters at the nominal values of $d=30$~mm, $h=8$~mm, $\delta=100$~\um, and \epsr=29.}
            \label{SimFig4}
        \end{figure}

        After identifying the resonance family and the role of the gap control, the remaining question is how the rest of the resonator design parameters can be used to localize plasma to different resonant locations at a desired frequency. For plasma generation, these parameters must control not only the resonant frequency, but also the energy storage and field concentration that determine whether breakdown can occur. To map these dependencies, diameter $d$, height $h$, gap $\delta$, and relative permittivity \epsr were swept independently in eigenmode simulations while the remaining parameters were held at their nominal values of $d=30$~mm, $h=8$~mm, $\delta=100$~\um, and \epsr=29 (Fig. \ref{SimFig4}). To show that this design approach is not limited to one field structure, the sweeps were performed for both the \TM{21} family and the higher order \TM{31} family, which has a different field symmetry and occurs at a higher frequency (Fig. \ref{SimFig4}~B).
    
        Increasing diameter, height, or permittivity shifted $f_0$ downward for both mode families (Fig. \ref{SimFig4}~C). A larger diameter or height lengthens the transverse and axial resonant paths, while a larger \epsr shortens the wavelength inside the dielectric. In each case, the same field structure occurs at a lower frequency, consistent with Eq.~\ref{eqn:bessel_TM}. The gap behaved differently because it controls inter-resonator coupling rather than the dimensions of either disk. The resonant frequency varied weakly below about $\delta=100$~\um and rose more steeply at larger gaps as the disks moved away from the monolithic limit discussed in Section \ref{theory:gap_control}. An interesting note is that peaks in Q suggest that there are optimal parameter values that enhance certain modes specifically. This may be of use when designing the system to preferentially activate one or more of the eigenmodes unevenly. As expected, the \TM{31} family remained above the \TM{21} family throughout the sweep, consistent with Eq. \ref{eqn:bessel_TM}, showing that different mode families occupy distinct but tunable frequency regions.
    
        These results show that resonator geometry and material properties can be designed to prescribe the resonant frequencies and available field enhancement regions in the gap. The dielectric dimensions, permittivity, gap spacing, and waveguide boundaries set the resonant modes and determine where and how the electric field is amplified. The gap further controls the amplification available to each mode, so for a fixed incident power, it determines which resonant states can ignite and which localized regions reach the breakdown threshold. Changing the excitation frequency then selects among those resonant regions, allowing plasma to form at different locations without physically reconfiguring the device.

\section{Methods} \label{methods}
    \subsection{Experimental Apparatus} \label{methods:setup}
    
\begin{figure}[b!]
            \centering
            \includegraphics[width=1\textwidth]{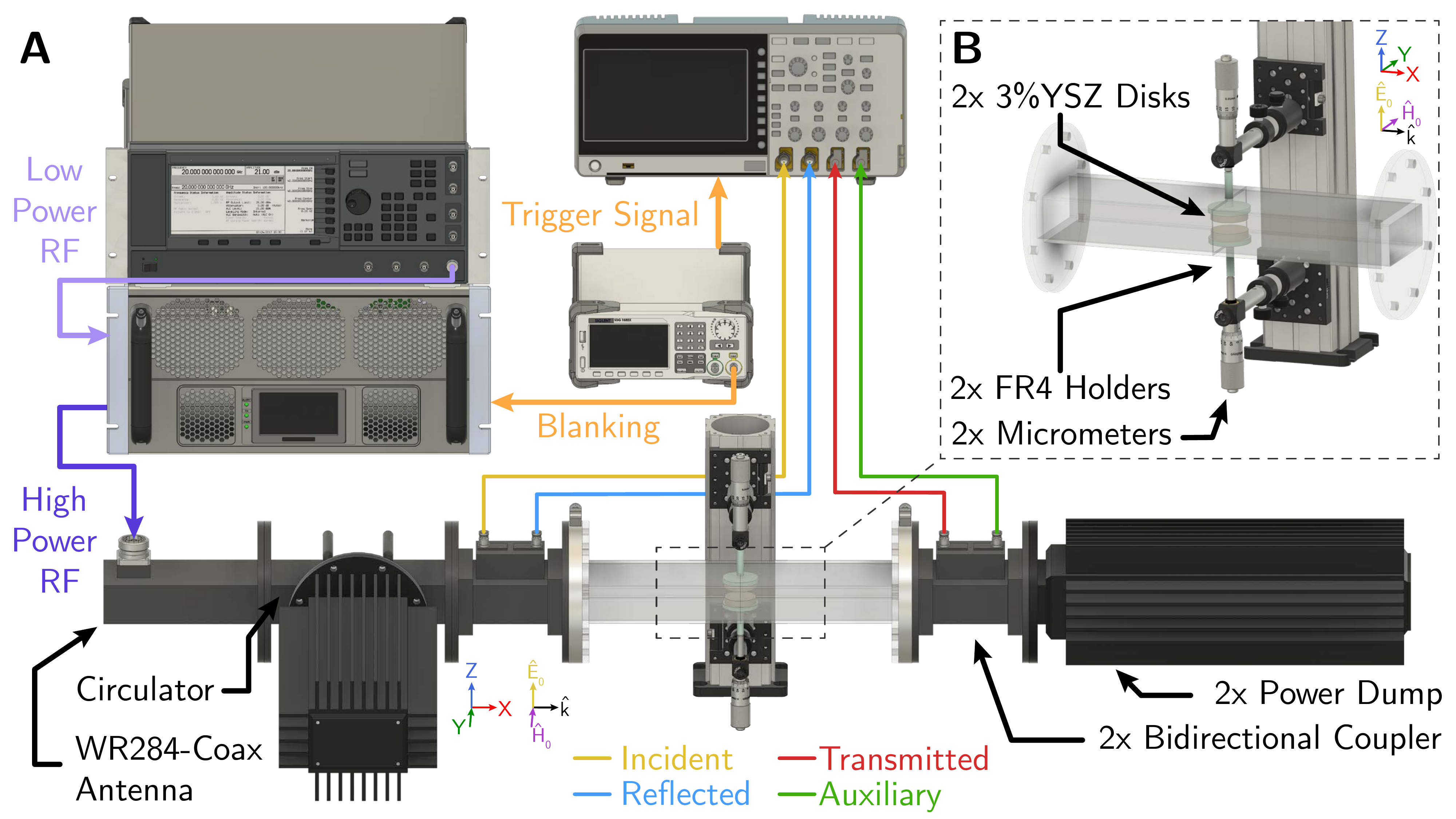}
            \vspace{-20pt}
            \caption{\textbf{A)} Experimental apparatus used to generate and characterize pulsed microwave breakdown between two dielectric resonators. A Keysight E8257D-1EU signal generator supplied the microwave signal to an EmpowerRF 2229-001 amplifier, gated by a Siglent SDG1032X waveform generator. A coaxial-to-WR-284 antenna launched the amplified signal into the waveguide as a \TE{10} mode, and a circulator redirected reflected power into a 50$\Omega$ matched load. Two bidirectional couplers around the test section sampled the incident, reflected, and transmitted microwave signals, which connected to four RFLambda RPDT0012GA power detectors feeding analog voltage into a Tektronix TBS 2104B oscilloscope. \textbf{B)} Two 3\% YSZ resonators were mounted in electrically insulating FR4 holders and positioned using external micrometer heads.}
            \label{MethodsFig1}
        \end{figure}

        The experimental setup delivered high-power, pulsed microwave signals with tunable excitation frequency, incident power, and pulse timing to a test section (Fig. \ref{MethodsFig1}). A programmable microwave signal generator supplied the incident waveform to a solid-state high-power amplifier, where an arbitrary waveform generator blanked the amplified signal into a square pulse. A coaxial-to-WR-284 antenna launched the amplified signal as a \TE{10}-polarized wave into the waveguide. A high-power circulator directed this wave toward the test section and routed reflected power into an impedance-matched load, protecting the source hardware from impedance mismatches in the waveguide and transmission line that would otherwise reflect large amounts of power back toward the amplifier. The test section held two dielectric resonators in fiberglass frames, suspended and separated by an adjustable gap. Bidirectional couplers on either side sampled the incident, reflected, and transmitted signals, while a downstream matched load absorbed transmitted power to prevent reflections back toward the test section (Fig. \ref{MethodsFig1}~A).
        
    \subsection{Microwave Signal Generation and Amplification} \label{methods:equipment}
    
        Independent control of frequency, output power, and pulse timing allowed the incident signal waveform to be tuned and shaped. A low-power programmable signal generator, a high-power amplifier, and a waveform generator were used to modify each quantity independently. A Keysight E8257D-1EU PSG analog signal generator produced the incident signal, tunable from 2.80 to 3.60~GHz with programmed output powers between -18 and -2~dBm. Its internal step and list sweeps executed the discrete frequency and power sequences described in each experimental procedure, with switching times under 11 ms between step states.
    
        An EmpowerRF Systems 2229-001 solid-state amplifier boosted the low-power signal from the Keysight PSG with a maximum gain of 64~dB and a peak pulsed output rating of 2.5~kW ($\sim$64~dBm). The amplifier operated in pulsed mode across pulse repetition frequencies (PRF) of 0.5 to 25 kHz and duty cycles (DC) up to 20\%, which together bounded the available pulse width (PW). Output rise and fall times were approximately 70~ns between the 10\% and 90\% levels.
        
        A Siglent SDG1032X waveform generator supplied the pulse and synchronization timing. The output channel provided the amplifier with a blanking signal, setting the PRF and DC, while a synchronized AUX output triggered the oscilloscopes and imaging systems so that electrical and optical measurements aligned with the microwave pulse. The measurements reported here used pulse repetition frequencies of 400--1000 Hz and DC values of 10--20\%, selected to produce pulse widths from 200--500~\us. Longer 500~\us pulses were used for single-pulse temporal measurements to let the plasma stabilize, while shorter pulse widths were used in modulation measurements to reduce the off time between powered regions and encourage plasma sustainment.
    
    \subsection{Waveguide Transmission Line} \label{methods:waveguides}
    
\begin{figure}[!b]
            \centering
            \includegraphics[width=1\textwidth]{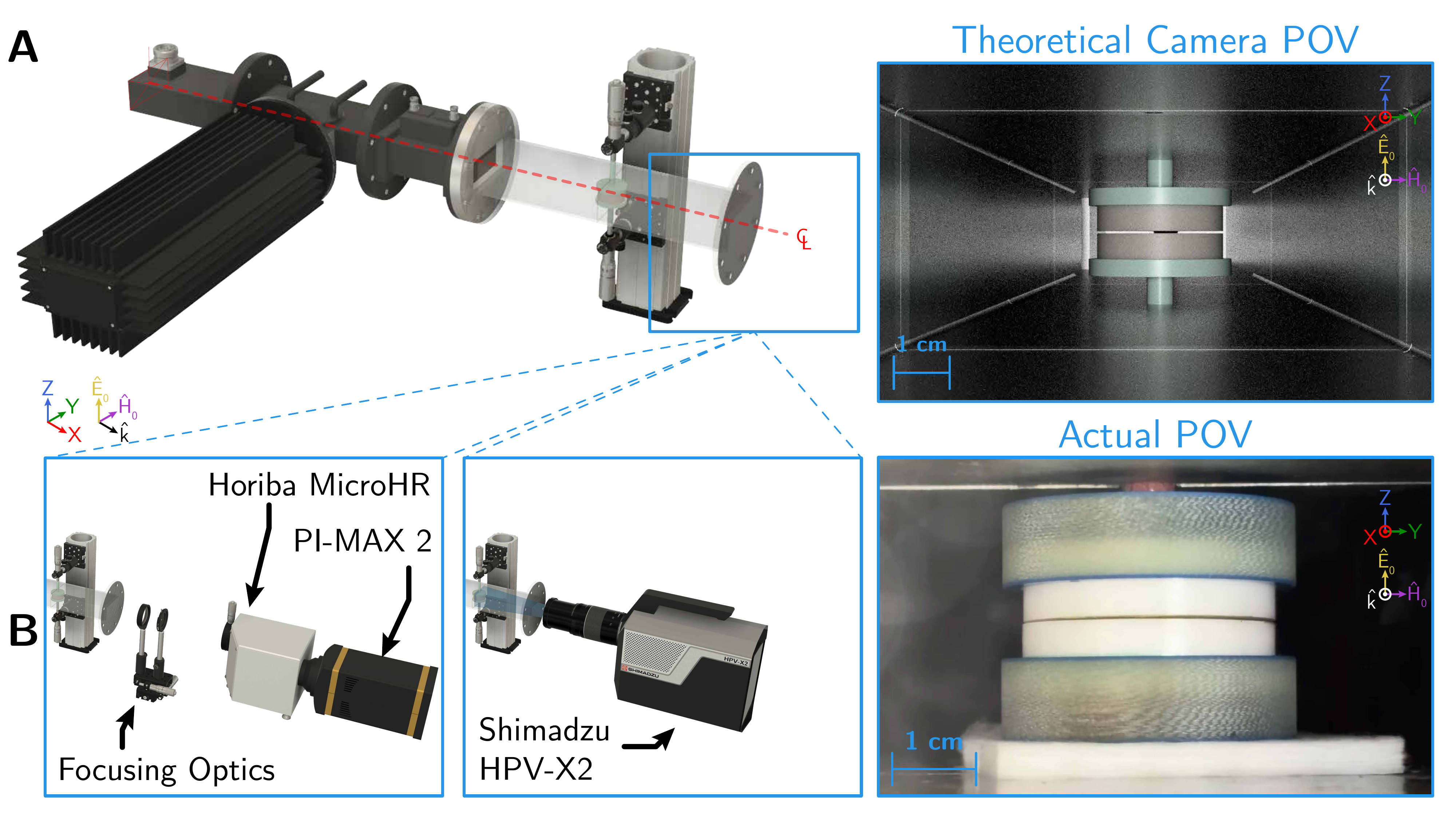}
            \vspace{-25pt}
            \caption{\textbf{A)} Experimental setup with reflective mesh grating replacing the transmitted section to allow for axial visibility into the gap region while confining the transmitted signal inside the test section. Sample images from CAD and experiment are shown to the side. \textbf{B)} Alternate configurations enabling either optical emission spectroscopy or high-speed imaging of the plasma created within the gap region.}
            \label{MethodsFig2}
        \end{figure}

        Waveguide components from CT Microwave were used as the transmission line of the experimental setup. A coaxial-to-waveguide antenna launched the amplified signal as the \TE{10} mode of the WR-284 waveguide through a 7/16 DIN coaxial connection. The 2.80--3.60~GHz experimental range fell within the 2.06--4.16~GHz single-mode band of a WR-284 waveguide. Downstream of the antenna, a three-port circulator, two bidirectional couplers, the dielectric-resonator test section, and impedance-matched loads were used (Fig. \ref{MethodsFig1}~A). The circulator directed incident power toward the test section and routed reflected power into a matched load to protect the antenna and amplifier circuits. A second impedance-matched load terminated the downstream end of the test section and absorbed transmitted power after it passed through the resonators (Fig. \ref{MethodsFig2}).
    
    \subsection{Microwave Power Diagnostics} \label{methods:power_io}
        
        The two bidirectional couplers sampled the incident, reflected, and transmitted microwave signals at the test section. Each coupled port had a nominal $-30$~dB coupling factor, sampling roughly one-thousandth of the propagating power. An additional fixed 20~dB attenuator was connected to each coupled port, providing approximately 50~dB of total attenuation before power detection. Four RFLambda RPDT0012GA coaxial power detectors then converted the incident, reflected, and transmitted signals into voltage outputs between 0 and 2048~mV. The analog output could then be correlated to power through a nonlinear calibration curve.
    
        A Tektronix TBS2104B oscilloscope recorded the detector voltages, triggered by the waveform generator synchronization signal or by the auxiliary trigger from the signal generator during programmed sweeps. Recorded timing signals, including the amplifier blanking and source-settled signals, verified temporal alignment among the source, amplifier, oscilloscopes, and optical diagnostics. Each measurement channel was calibrated independently over the experimental frequency and power ranges, accounting for the coupling factor, fixed attenuation, cable losses, and detector response to convert from mV to dBm. The source-settled and blanking signals separated each acquisition into its programmed step conditions. Data recorded during source transitions were excluded, and the remaining record was sorted into sweep-step bins and individual pulses referenced to the rising edge of each powered pulse. Pulse-resolved quantities were computed for each valid pulse before ensemble statistics were evaluated.
    
    \subsection{Down Axis Imaging} \label{methods:axial_imaging}
    
        For plasma imaging, a microwave-reflective, optically transmissive grating replaced the downstream matched load, providing axial optical access to the inter-resonator gap while acting as a reflecting microwave boundary (Fig. \ref{MethodsFig2}~A). A Shimadzu HPV-X2 high-speed camera viewed the gap through this grating at frame rates up to 2~M\fps, giving 400~ns exposure with 500~ns inter-frame spacing at the maximum rate. A synchronized 5~V TTL signal triggered the camera, and a programmable delay set the acquisition window relative to the start of the powered pulse. An Apple 48~MP Dual Fusion camera recorded longer time-integrated color video through the same axial path at 4K resolution and 60~\fps, giving a qualitative visualization of the plasma structure across pulse trains and during modulation sweeps. Since each frame integrated emission over a far longer interval than the HPV-X2 exposure, these videos were not used to resolve intra-pulse dynamics.
    
    \subsection{Dielectric Resonator Loaded Test Section} \label{methods:test_section}
    
        The dielectric disks were pressed and sintered from 3\% yttria-stabilized zirconia and machined into cylindrical resonators approximately 30~mm in diameter and 8~mm thick. Dense YSZ ceramics exhibit relative permittivities of approximately \epsr=21--32 and microwave loss tangents on the order of $\tan\delta=1\times10^{-3}$--$6\times10^{-3}$ at room temperature, depending on yttria concentration, density, microstructure, temperature, and frequency \cite{Park2022SynthesisProperties, Oh2019MicrowaveJetting}. The resonators used here had an estimated relative permittivity of approximately \epsr=25.4 over 2.7--3.5~GHz (Fig. \ref{MethodsFig3}).
    
        The two resonators were positioned in the waveguide to form a coupled unit cell in which the adjustable spacing between the opposing circular faces allowed the external resonant fields to overlap and concentrate electric field in the gap. The resonator axes were aligned with the incident electric field polarization $\hat{E}$, which lies along the short axis of the waveguide, while the wave propagated along the waveguide axis $\hat{k}$. In this orientation, the incident microwave field propagated across the resonator diameters rather than along their axes or through their thicknesses (Fig. \ref{MethodsFig2}). This orientation concentrated the resonant field within the inter-resonator gap, where breakdown occurred, while permitting axial imaging through the downstream grating. Custom holders machined from 0.25~in thick flame-retardant Garolite G-10/FR4 constrained the radial and angular positions, isolated the resonators from the conductive waveguide walls, and allowed independent translation using two Mitutoyo 150-801 micrometer heads. Each micrometer provided 0 to 25~mm of travel with 1~\um graduations and approximately $\pm2$~\um instrumental error. The gap was set from the relative micrometer displacement after establishing a zero-gap reference, with additional uncertainty from mechanical backlash, holder compliance, and face alignment.

    \subsection{Temperature Induced Resonance Shifts} \label{methods:temp_shift}

\begin{figure}[!t]
            \centering
            \includegraphics[width=1\textwidth]{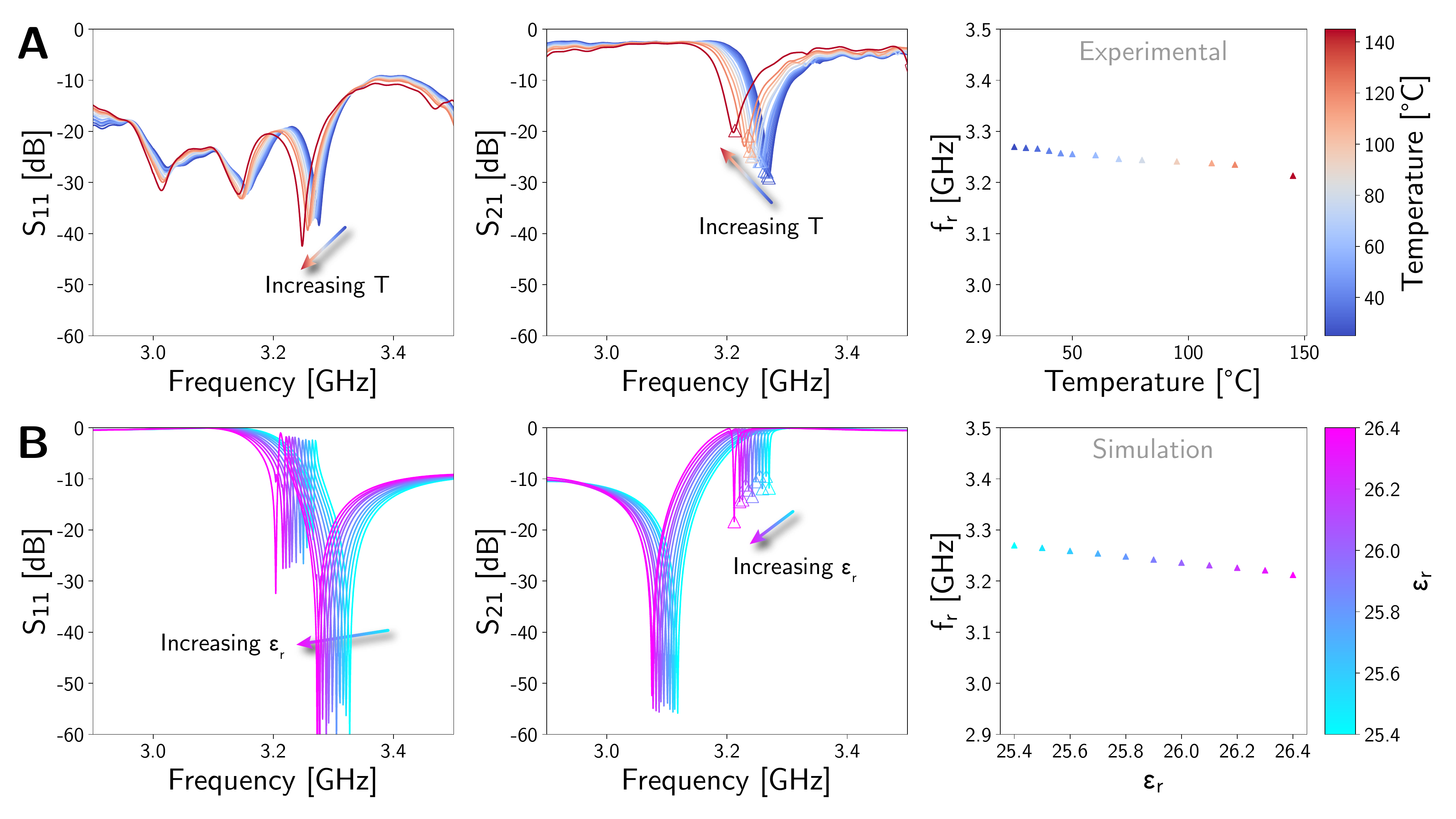}
            \vspace{-25pt}
            \caption{Comparison of experimental \textbf{A)} and simulated \textbf{B)} $S_{11}$ and $S_{21}$ responses demonstrating the effects of resonator temperature and relative permittivity on the resonant-frequency response. Experimental spectra were measured over a resonator-temperature range of 35--140~\degC, with increasing temperature indicated from blue to red. Simulated spectra were calculated by varying the resonator relative permittivity from \epsr=25.4 to 26.4, with increasing permittivity indicated from cyan to magenta. Arrows indicate the direction of the frequency shift with increasing temperature and relative permittivity.}
             \label{MethodsFig3}
        \end{figure}

        Experiments were performed to identify how dielectric heating shifted the resonant conditions of the coupled assembly. This characterization was necessary because the plasma-localization strategy relies on exciting selected resonant modes, and even modest frequency drift can change the local field enhancement, apparent breakdown threshold, and plasma structure. In 3\% YSZ, the real and imaginary components of the complex permittivity depend on temperature, so dielectric heating provides a direct mechanism for thermally induced resonance drift. The temperature-dependent measurements were therefore used to separate intentional frequency tuning from thermally induced shifts in the resonant response (Fig. \ref{MethodsFig3}).
    
        A Siglent SNA5002A two-port vector network analyzer measured the low-power $S_{11}$ and $S_{21}$ response over the 2.80--3.50~GHz range using an output power of 10~dBm (0.01~W). For these measurements, the source, amplifier, and high-power delivery components were replaced by the VNA through two N-type-to-WR-284 transitions. The disks were heated to approximately 175~\degC by tuning the microwave source to a strongly absorptive resonant condition below the plasma ignition threshold. The heating signal was cycled to allow heat to spread through the disks, and the resonator temperature was measured at 1~min intervals using an infrared thermal gun and a contact thermocouple. Once the measured surface temperature was approximately uniform across multiple resonator locations, the high-power connections were removed, and the VNA was attached. The VNA then collected low-power spectral sweeps as the disks cooled toward room temperature.
    
        Experimental spectra collected during cooling from 140~\degC to 35~\degC were compared with simulated spectra calculated over a relative permittivity range of \epsr=25.4 to 26.4. This comparison gave a room-temperature permittivity estimate of approximately \epsr=25.4 for the resonators used in this study. Aligning the experimental and simulated spectra showed that increasing either resonator temperature or relative permittivity shifted the resonant features downward in frequency (Fig. \ref{MethodsFig3}). Over 35 to 140~\degC, the primary $S_{11}$ and $S_{21}$ features used for tracking shifted from approximately 3.27 to 3.21~GHz. This corresponded to an approximately linear shift of about 60~MHz, or roughly 0.5--0.6~MHz per \degC. The simulations varied only the real permittivity of the resonators. They were therefore used to interpret the temperature-induced frequency shift, while changes in peak depth and bandwidth were attributed to temperature-dependent loss and coupling effects that were not modeled explicitly.
    
        The thermal drift measured here was important for interpreting the plasma experiments because dielectric heating changes the resonator permittivity and therefore its refractive index. This shifts the resonant band, so a fixed excitation frequency can move closer to, between multiple, or farther from resonant modes as the disks heat. For the single pulse and modulation measurements used in subsequent sections, the pulse sequences were short enough that heating was treated as negligible between individual steps. Cooling between runs limited large temperature offsets between datasets. Over longer operation, however, accumulated heating could shift neighboring resonances into or out of the excitation band and change the plasma structure, ignition threshold, or sustainment behavior.

\section{Results and Discussion} \label{results}

    \subsection{Eigenmode Selected Microwave Breakdown Patterns} \label{results:eigenmode_breakdown_patterns}

        Experiments were performed to test whether plasmas could be formed at the resonant subwavelength voids that were predicted by electromagnetic simulations. The coupled resonator simulations predicted a closely spaced sequence of resonant responses, each with a distinct electric field distribution in the inter-resonator gap and a corresponding absorption feature in the frequency response. 
    
\begin{figure}[t!]
            \centering
            \includegraphics[width=1\textwidth]{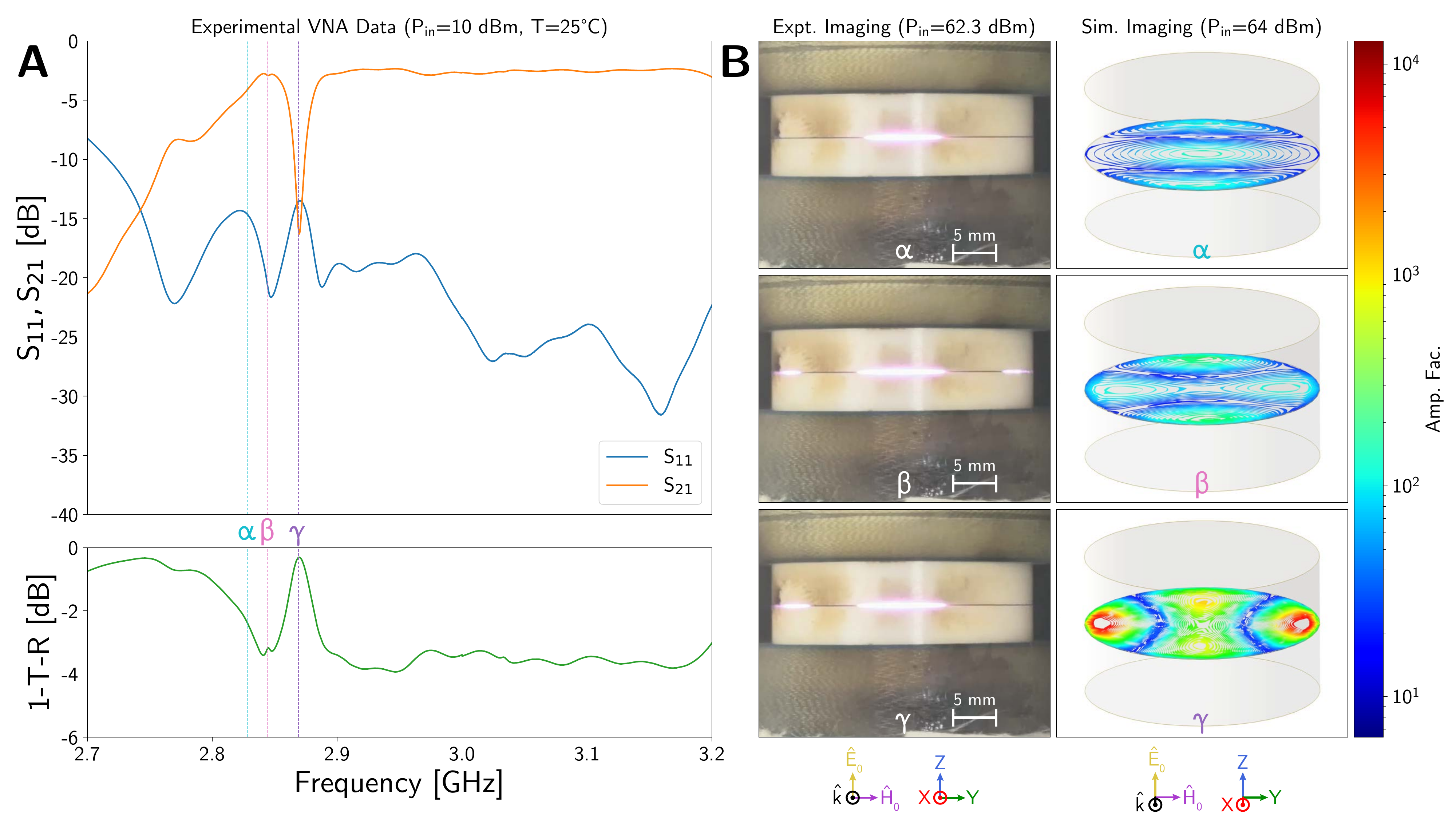}
            \vspace{-25pt}
            \caption{\textbf{A)} low-power VNA measurement of $S_{11}$ and $S_{21}$ at an inter-resonator gap of approximately 200~\um. The reflected, transmitted, and absorbed power fractions identify the resonant frequency range used to select the \modealpha, \modebeta, and \modegamma excitation frequencies. \textbf{B)} Plasma distributions produced in separate high-power measurements at the selected excitation frequencies compared with simulated electric field distributions of the corresponding coupled resonant modes.}
            \label{ResultsFig1}
        \end{figure}

        The dielectric resonator assembly was characterized with a low-power VNA sweep first at an incident power of 10~dBm to validate these predictions. During the sweep, the resonators were near room temperature, the waveguide contained room temperature atmospheric pressure air, and the inter-resonator gap was fixed at approximately 200~\um. The measured $S_{11}$ and $S_{21}$ spectra covered the frequency range containing the predicted resonant features. The absorbed power fraction was calculated from these scattering parameters during post-processing, and the resonant features were labeled \modealpha, \modebeta, and \modegamma (Fig. \ref{ResultsFig1}~A).

        The low-power spectra identified a closely spaced set of absorption features between approximately 2.800 and 2.900~GHz. The clearest correspondence with the simulated spectra occurred for the \modegamma condition at 2.869~GHz, where the sharpest absorption peak appeared with a decrease in transmission ($S_{21}$) and an increase in reflection ($S_{11}$). The \modebeta condition was identified at 2.844~GHz, where a more subtle absorption peak occurred with a corresponding dip in reflection while transmission remained approximately constant. The \modealpha condition was selected at 2.828~GHz within a broader absorption feature extending from approximately 2.76 to 2.82~GHz. Simulations indicated that this region also contained a secondary \TM{012} response that absorbed power but did not strongly amplify the gap field. Its field varied mainly along the disk height, so it did not constructively concentrate between the opposing dielectric surfaces. This overlap partially obscured the \HE{21}{-} dominated feature in the low-power spectrum, but the high-power plasma distribution remained consistent with the \modealpha field localization pattern.
        
        The resonant frequencies identified from the low-power response provided the operating points for high-power measurements of plasma localization. For these measurements, the VNA was replaced with the signal generator and amplifier configuration, and the downstream VNA transition was replaced with the mesh grating for down-axis imaging. The dielectric resonator assembly was operated under the same nominal thermal and geometric conditions used in the low-power sweep. Plasma images were acquired using a burst train of 10 pulses at 400~Hz pulse repetition frequency, 20\% duty cycle, and 500~\us pulse width, with an incident power of approximately 62.3~dBm, or 1.7~kW. The excitation frequencies were 2.828, 2.844, and 2.869~GHz for \modealpha, \modebeta, and \modegamma, respectively. A 60~Hz down-axis video recorded time-integrated broadband optical emission from the air plasma, and the resulting images were compared with the simulated field structures expected for each resonant response (Fig. \ref{ResultsFig1}~B).
    
        The down-axis plasma images showed that breakdown remained confined to the inter-resonator gap and aligned with the simulated electric field patterns for all three resonant conditions. Although axial imaging did not resolve the full three-dimensional plasma distribution, the projected emission patterns still allowed direct comparison with the simulated field patterns for each mode. For \modealpha at 2.828~GHz, the plasma appeared near the center of the disks. For \modebeta at 2.844~GHz, the emission is separated into three apparent regions, consistent with the predicted four-point field structure when projected along the imaging axis. In this projection, the front and rear high-field locations could appear aligned as a single central emission region. For \modegamma at 2.869~GHz, the plasma shifted toward one side of the gap while retaining emission near the center. The images showed that the breakdown was localized to regions where the dielectric resonance enhanced the electric field above threshold, while lower field regions remained unignited. These observations provide the central spatial control result. The dielectric resonances determined where breakdown occurred, and the excitation frequency selected which resonant field structure produced the plasma.
    
        The \modealpha, \modebeta, and \modegamma patterns arise from the cylindrical disk geometry used here, but the mechanism is general. Dielectric geometry, material properties, gap spacing, and waveguide boundaries determine which resonant modes are excited and how their fields are concentrated in the gap between surfaces. Other resonator shapes or coupled dielectric designs would therefore produce different field maps while using the same gap-based mechanism for localized breakdown near a target surface.

    \subsection{Frequency Selection of Resonant Plasma Distributions} \label{results:frequency_control}

        Frequency modulation experiments were performed to demonstrate that the microwave excitation frequency could serve as an active control parameter to switch the location of plasma in time. Section \ref{results:eigenmode_breakdown_patterns} showed that changing the excitation frequency selected different resonant field structures and, in turn, different breakdown locations within the coupled dielectric assembly. Here, the excitation frequency was changed between individual microwave pulses so that successive pulses could excite different resonant modes within one pulse train. This pulse-to-pulse frequency control tested whether the plasma distribution could be reconfigured programmatically by selecting among the \modealpha, \modebeta, and \modegamma resonant states.
    
        Two experiments were used to demonstrate this control. First, low-power and high-power discrete frequency sweeps were compared to identify the resonant band where plasma ignition occurred and to determine how breakdown changed the reflected, transmitted, and absorbed power fractions relative to the low-power response. Second, targeted frequency steps were applied at low and high-power to test whether the plasma distribution could be switched between the \modealpha, \modebeta, and \modegamma states in quick succession.

       \subsubsection{Plasma Loaded Response During Discrete Frequency Sweeps} \label{results:frequency_control:frequency_sweep}

            The same resonant frequency conditions were measured at low and high incident power so that the microwave response before and after breakdown could be compared directly. The low-power sweep remained below breakdown and measured the resonator response without plasma, while the high-power sweep produced plasma when the driven field enhancement exceeded the breakdown threshold. Comparing the two sweeps showed how plasma formation modified the reflected, transmitted, and absorbed power fractions of the same dielectric resonator assembly.
        
\begin{figure}[t!]
                \centering
                \includegraphics[width=1\textwidth]{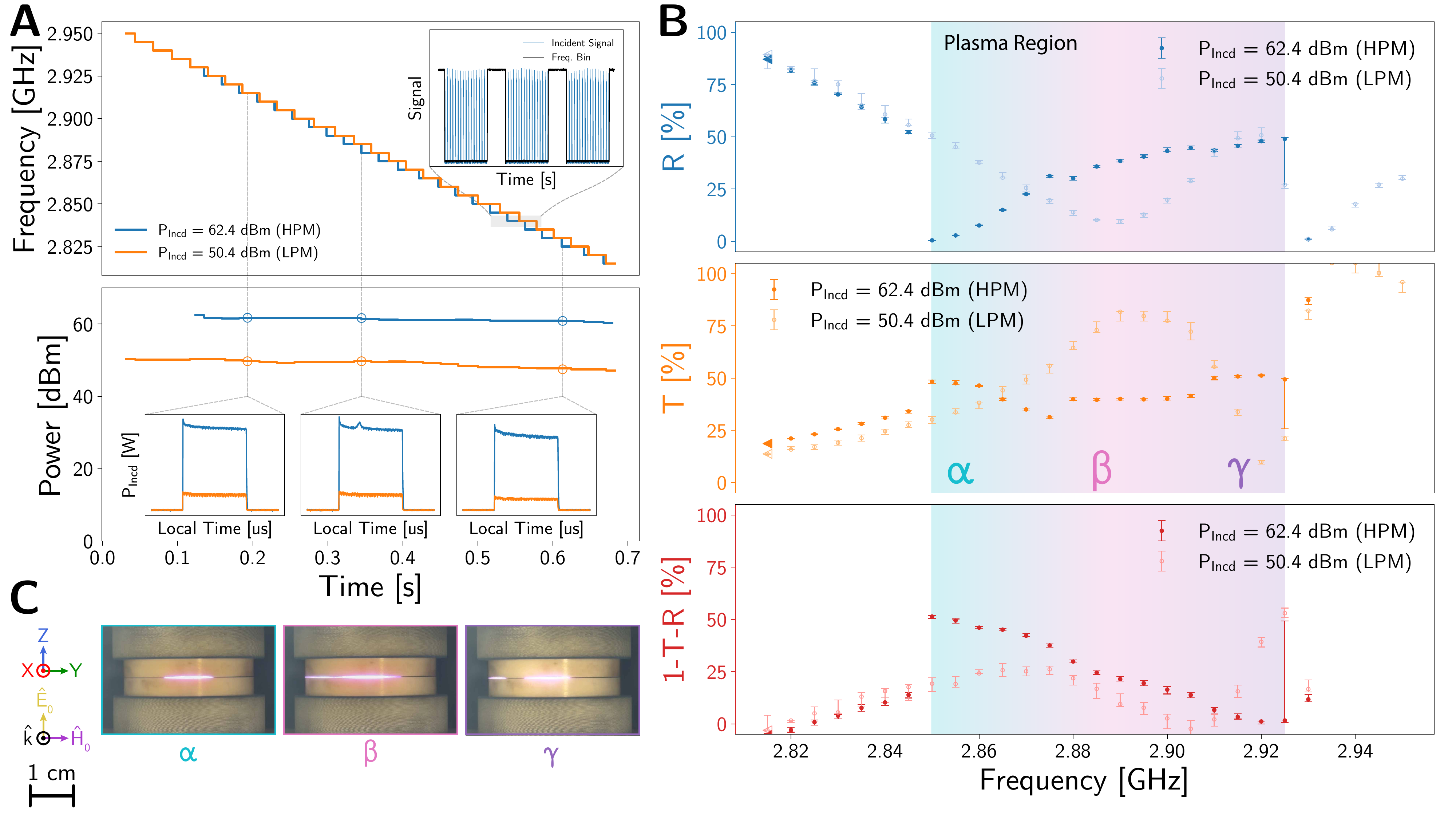}
                \vspace{-25pt}
                \caption{\textbf{A)} Commanded incident signal frequency and power during low-power and high-power discrete frequency sweeps. Power varied slightly across the sweep because of the frequency response of the signal generator and amplifier system, but remained approximately constant within each sweep. \textbf{B)} Reflected, transmitted, and absorbed fractions of the incident microwave power were evaluated at a fixed local time within each pulse. Error bars show the pulse-to-pulse variation within each frequency bin. The shaded region identifies the frequency band in which plasma was observed during the high-power sweep. \textbf{C)} Time-synchronized images acquired during the high-power sweep show that the plasma distribution changed through the \modealpha, \modebeta, and \modegamma patterns as the excitation frequency was varied.}
                \label{ResultsFig2}
            \end{figure}

            The inter-resonator gap was fixed at approximately 200~\um. The low-power sweep used an incident power of approximately 50.4~dBm, or 109.6~W, and the high-power sweep used approximately 62.4~dBm, or 1737.8~W. The excitation frequency was stepped from 2.95 to 2.815~GHz in 5~MHz increments, with a dwell time of 30~ms at each frequency. The full frequency sweep lasted approximately 2~s. The blanking signal applied to the amplifier produced microwave pulses at 1000~Hz PRF and 20\% DC, corresponding to a 200~\us PW. The reflected, transmitted, and absorbed fractions were evaluated at a fixed local time within each pulse, $T+175$~\us relative to the rising edge. The median response was plotted for each frequency bin, with error bars spanning the 25th to 75th percentiles (Fig. \ref{ResultsFig2}).
    
            The low-power sweep reproduced the principal resonant features observed in the electromagnetic simulations and VNA measurements, confirming that the signal generator sweep excited the same resonance family identified previously. The high-power sweep showed that plasma formation was confined to a finite resonant band from approximately 2.85 to 2.925~GHz (Fig. \ref{ResultsFig2}~B). Outside this band, the low-power and high-power sweeps produced similar reflected, transmitted, and absorbed fractions. Within this band, visible plasma appeared, and the high-power response departed from the low-power baseline. This comparison shows that high incident power alone did not produce breakdown. Breakdown occurred only when the excitation frequency drove a resonant field structure with sufficient gap field enhancement.
            
            Within the resonant breakdown band, the high-power response diverged from the low-power baseline, showing that plasma formation changed the microwave response of the coupled resonator assembly depending on the driven mode. Near \modealpha, breakdown increased absorption and reduced reflection; near \modebeta, it increased reflection and reduced transmission; and near \modegamma, it suppressed the sharp low-power absorption feature while substantially changing the reflected and transmitted fractions. These mode-dependent changes show that the localized plasma did not simply add broadband loss to the original resonator spectrum. Instead, the plasma, dielectric resonators, and waveguide formed a plasma-loaded resonant structure whose scattering response depended on where the plasma formed relative to the modal electric field. This interpretation is consistent with waveguide simulations showing that plasma located near a modal electric field maximum can have a much stronger effect on transmitted power than plasma located in weaker field regions \cite{Qu2017PropertiesWaves}. The corresponding images remained consistent with the confined \modealpha, \modebeta, and \modegamma plasma distributions identified in the previous section (Fig. \ref{ResultsFig1}~B).
           
            These results show that the observed responses are not isolated conditions, but frequency-selected eigenmode superpositions within a coupled resonant spectrum. The discrete sweep, therefore, demonstrated spatial localization by showing that different excitation frequencies produced distinct plasma distributions in separate regions of the resonator gap. The next section tests this control more directly by switching the excitation frequency between selected resonant states within a programmed pulse sequence.

        \subsubsection{Rapid Resonant Mode Selection with Targeted Frequency Steps} \label{results:frequency_control:three_step}

            After the discrete sweep identified the resonant breakdown band, a targeted frequency step experiment tested whether selected plasma states could be formed on demand. Instead of sweeping continuously through the resonance band, the incident frequency was switched between three preset conditions corresponding to the dominant \modealpha, \modebeta, and \modegamma responses. This tested whether the plasma distribution could be rapidly reconfigured by selecting the excitation frequency alone, without passing through the intermediate resonant states.
            
\begin{figure}[t!]
                \centering
                \includegraphics[width=1\textwidth]{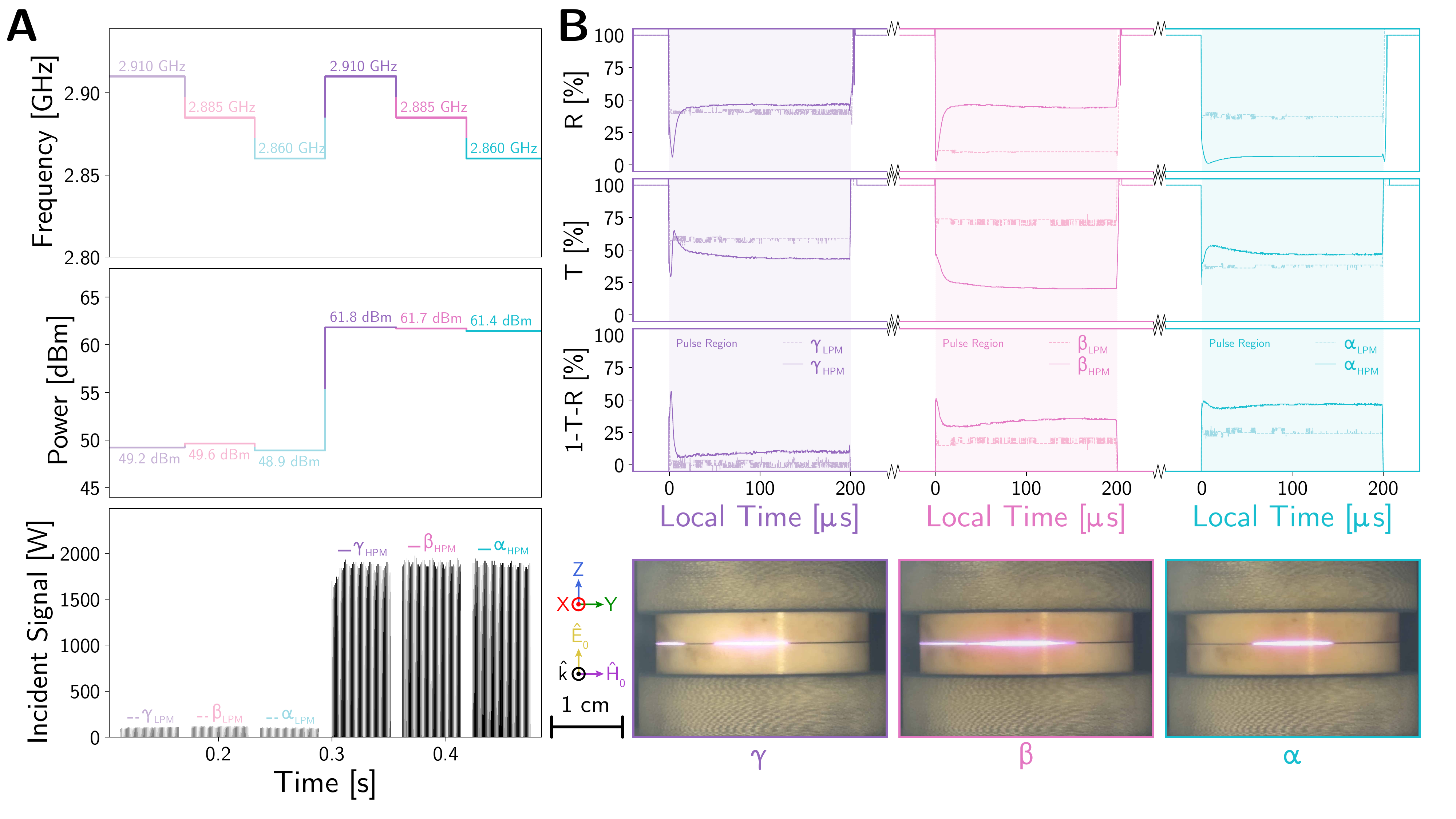}
                \vspace{-25pt}
                \caption{\textbf{A)} Commanded incident signal frequency and power during a six-condition sequence combining three selected excitation frequencies with low-power and high-power states. \textbf{B)} Reflected, transmitted, and absorbed fractions of the incident microwave power for the median pulse within each frequency and power condition. \textbf{C)} Images acquired during the high-power frequency steps show the plasma distributions associated with the selected \modealpha, \modebeta, and \modegamma conditions.}
                \label{ResultsFig3}
            \end{figure}

            Three excitation frequencies of 2.86, 2.885, and 2.910~GHz were selected to target the dominant \modealpha, \modebeta, and \modegamma structures, respectively. A six-condition sequence combined these three frequencies with low-power and high-power states (Fig. \ref{ResultsFig3}). The low-power conditions, approximately 48--49~dBm, measured the response before breakdown at each frequency, while the high-power conditions, approximately 61.5--62~dBm, produced plasma. The same 1000~Hz PRF, 20\% DC, and 200~\us PW used in the preceding sweep were retained, giving approximately 50 pulses per step. The dwell time was increased to 50~ms per condition so that each selected plasma distribution could develop before the next step. The full six-condition sequence required at most approximately 350~ms, compared with nearly 2~s for the full discrete frequency sweep in Section \ref{results:frequency_control:frequency_sweep}. Each plotted trace represents the median response from all valid pulses within the corresponding frequency and power condition, so the traces describe the repeatable response of each selected state rather than a single pulse.
            
            The targeted frequency steps reproduced the plasma-loaded response expected for each resonant state. The \modealpha, \modebeta, and \modegamma conditions produced the same mode-dependent changes in reflected, transmitted, and absorbed power observed in the discrete sweep, while the time-integrated images confirmed that each frequency selected the corresponding plasma distribution (Fig. \ref{ResultsFig3}~C).

            This experiment demonstrated programmed selection of resonant plasma states through pulse-to-pulse frequency modulation. The discrete sweep showed that plasma localization shifted across the resonant band, while the targeted sequence showed that selected states could be accessed directly without scanning intermediate frequencies. By modulating only the incident frequency, the plasma could be extinguished by tuning off resonance and reignited by tuning back onto any resonant state. The resonator geometry, therefore, defined the set of available plasma states, while the excitation frequency rapidly selected which state was activated.

    \subsection{Amplitude Control of Plasma Sustainment and Ignition} \label{results:amplitude_control}
    
        The preceding experiments showed that excitation frequency selects the resonant field structure and therefore the preferred location for plasma formation near the dielectric surfaces. The next step was to determine how the remaining pulse parameters control access to that selected state. In resonant discharges, microwave pulse shaping involves controlling the excitation frequency, peak power, and average power. Excitation frequency selects which resonant mode is driven, peak power determines whether the local field exceeds the breakdown threshold, and average power influences dielectric heating and the longer time discharge state. These quantities are coupled because each mode has a different quality factor and field amplification, so stronger resonant enhancement should reduce the incident peak power that is required for ignition.

        Power was therefore varied at fixed \modealpha, \modebeta, and \modegamma excitation frequencies to identify how each selected resonant state responds to changes in incident amplitude. The decreasing power sweeps were used to measure the sustainment limit after ignition, while a bidirectional sweep at the \modegamma frequency was used to compare the incident power required to initiate breakdown with the power required to maintain it.

        \subsubsection{Power Modulation of Each Mode} \label{results:amplitude_control:power_sweeps}
        
\begin{figure}[t!]
                \centering
                \includegraphics[width=1.0\textwidth]{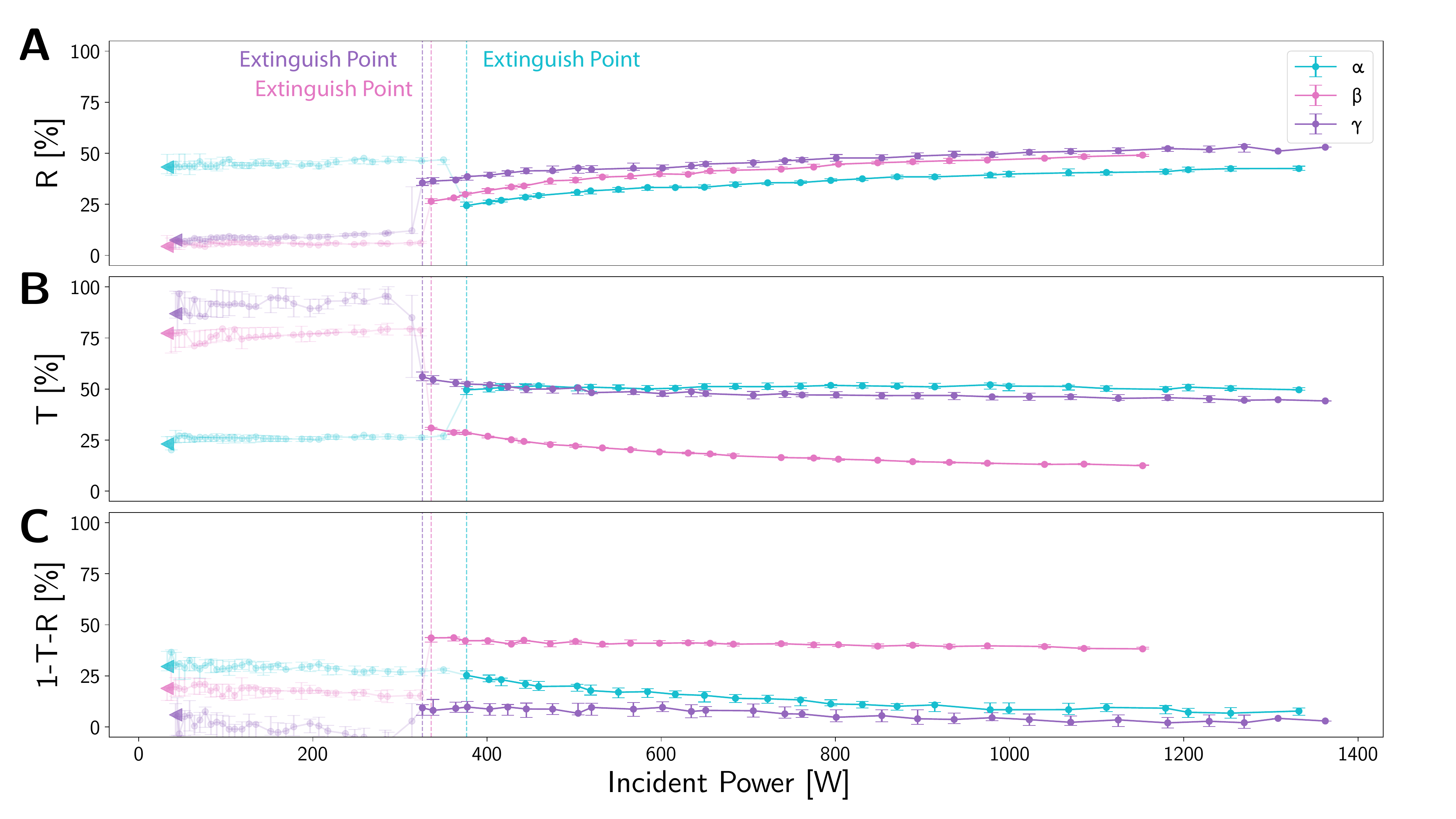}
                \vspace{-15pt}
                \caption{\textbf{A)} Reflected, \textbf{B)} transmitted, and \textbf{C)} absorbed fractions of the incident microwave power during decreasing power sweeps at the selected \modealpha, \modebeta, and \modegamma excitation frequencies. Each sweep began from a plasma on condition and was reduced in power until the discharge extinguished. The \modegamma plasma remained sustained to the lowest incident power, indicating that this resonant condition provided the strongest sustainment under the tested conditions.}
                \label{ResultsFig4}
            \end{figure}

            The incident power was reduced through a stepwise sweep to compare how the \modealpha, \modebeta, and \modegamma states sustained breakdown through their mode-dependent field amplification. The excitation frequencies were held fixed at 2.86, 2.885, and 2.910~GHz, respectively, matching the resonant states used in the rapid frequency modulation measurements (Fig. \ref{ResultsFig3}). Each sweep began at approximately 1400~W to establish the plasma, after which the incident power was reduced stepwise until the discharge extinguished (Fig. \ref{ResultsFig4}). Since the three modes have different quality factors, coupling strengths, and gap field enhancements, they were not expected to remain ignited to the same incident power. The extinction power, therefore, provided a measure of how effectively each selected resonant state amplified and sustained the local field needed for breakdown.

            Each power step was held for 30~ms using a 1000~Hz pulse repetition frequency and 20\% duty cycle, corresponding to approximately 30 pulses per bin with a 200~\us pulse width. Reflected, transmitted, and absorbed power fractions were extracted at $T+175$~\us, where $T+0$ denotes the start of the powered pulse. The median response was plotted for each power bin, with error bars spanning the 25th to 75th percentiles. Larger inter-quartile ranges near extinction indicate delayed or intermittent plasma loss within the bin.

            Each resonant condition showed an abrupt transition when the discharge extinguished, with the microwave response returning toward the low-power behavior identified in Section \ref{results:frequency_control} rather than drifting gradually (Fig. \ref{ResultsFig4}). The extinction power depended on the driven resonant state: \modealpha extinguished at approximately 380~W, or 55.8~dBm, while \modebeta remained sustained to approximately 330~W, or 55.2~dBm, and \modegamma persisted to approximately 316~W, or 55.0~dBm. This ordering shows that the same incident power does not produce the same plasma strength for every mode. Instead, the sustainment limit depends on the quality factor, coupling strength, and field amplification of the selected resonance, with more strongly amplified modes maintaining breakdown at lower incident power.
            
            These sweeps show that control over peak power acts through the resonant amplification of each selected state rather than uniformly raising or lowering all discharges. This motivates pulse shaping in which a high peak power pulse first establishes plasma-loaded coupling, followed by lower incident power to sustain the selected mode. The following hysteresis measurement tests this separation by comparing ignition and extinction thresholds for the strongest sustainment condition.

        \subsubsection{Hysteresis Between Plasma Ignition and Extinction} \label{results:amplitude_control:plasma_hysteresis}

            Stepwise power reduction measurements established how far each selected plasma state could remain sustained once the discharge was formed. A bidirectional power sweep was then performed to separate the incident peak power required to initiate breakdown from the lower power required to maintain the established plasma state. The upward sweep measured the power needed to initiate breakdown from a plasma-free state, while the downward sweep tested whether the established plasma could remain ignited at reduced incident power. This distinction is important for pulse shaping because a high-power portion of the pulse can initiate the discharge, after which the power can be reduced to lower the average power deposited into the dielectrics while maintaining the plasma.

\begin{figure}[t!]
                \centering
                \includegraphics[width=1\textwidth]{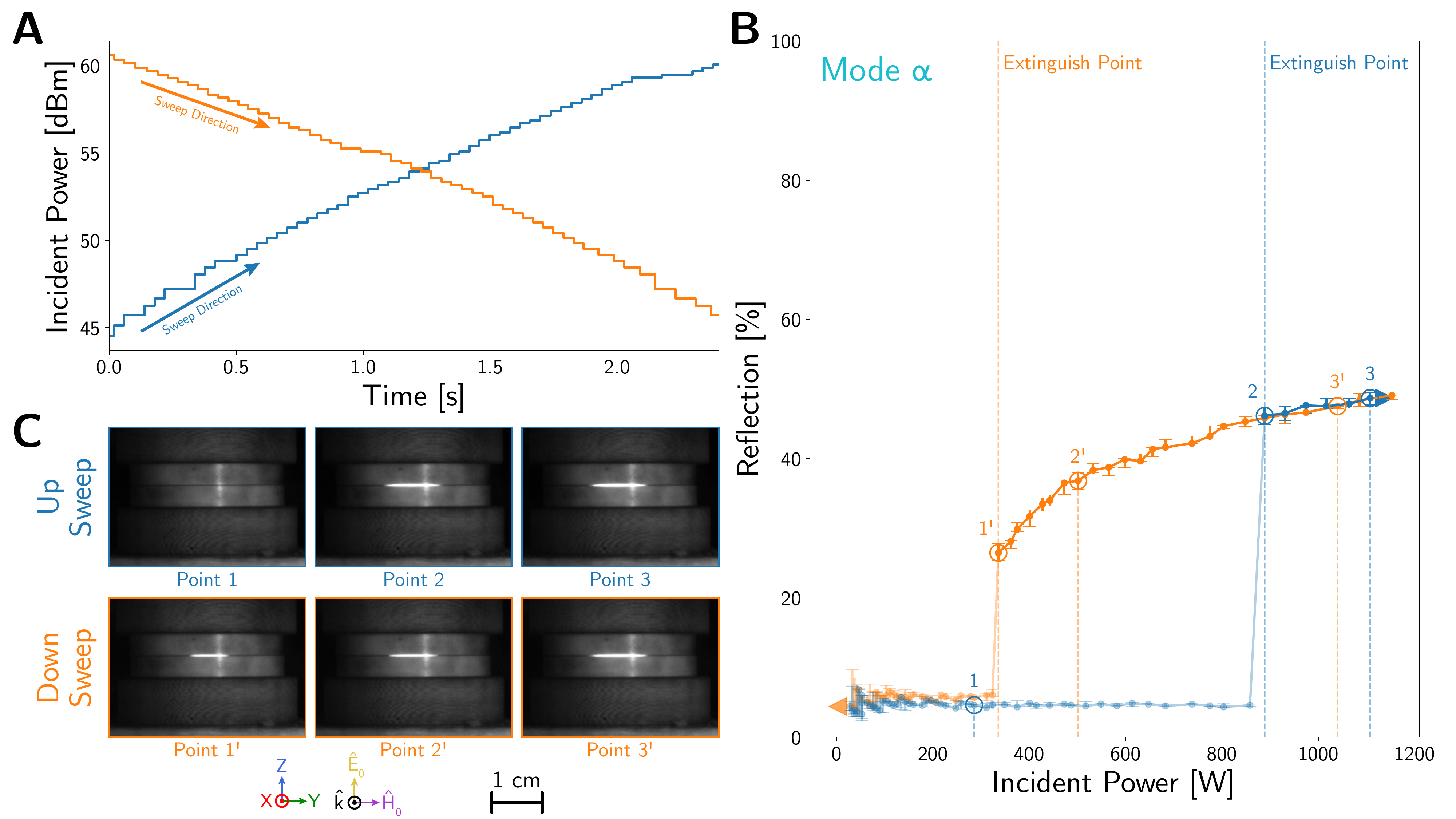}
                \vspace{-20pt}
                \caption{Hysteresis between the ignition and extinction thresholds of the \modealpha plasma. During the increasing power sweep, plasma ignition occurs at approximately 800--850~W. During the decreasing power sweep, the established plasma remains sustained until the incident power falls to approximately 300--350~W. Time-integrated images with 1 ms exposure (100 fps) at points 1--3 and 1'--3' show the plasma state along the increasing and decreasing power branches, respectively.}
                \label{ResultsFig5}
            \end{figure}

            The \modealpha condition was selected for this comparison because it produced the most visually distinct plasma structure when viewed along the waveguide axis. This made it the clearest case for comparing the plasma-free, ignited, and sustained states using time-integrated imaging. The excitation frequency was fixed at 2.860~GHz, and the incident power was first increased from approximately 45~dBm to 61.5~dBm before being decreased through the same values in reverse order. All other parameters, including inter-resonator gap, dwell time, PRF, duty cycle, pulse width, pulse processing, and statistical analysis, were kept the same as in the preceding decreasing power sweeps. The reflected power fraction and time-integrated plasma images were recorded along both trajectories to identify the ignition and extinction thresholds (Fig. \ref{ResultsFig5}). Time-integrated images were acquired with the high-speed camera during both the increasing and decreasing power sweeps at 100~fps with a 1~ms exposure, allowing changes in plasma structure to be synchronized with the microwave response.
            
            The bidirectional sweep separated the power required for ignition from the power required for sustainment. During the increasing power sweep, the \modealpha field structure did not ignite until the incident power reached approximately 800--850~W, where the reflected fraction changed abruptly, and optical emission appeared in the gap. During the decreasing power sweep, the established plasma remained sustained until the incident power fell to approximately 300--350~W. This hysteresis shows that the resonant plasma state could be maintained at substantially lower power than was required to first ignite it.

            Time-integrated imaging confirmed that the plasma state depended on both incident power and prior ignition history. At the same incident power, the increasing branch remained below the breakdown threshold while the decreasing branch remained ignited. As the incident power decreased, the emission region contracted, showing that only the most strongly amplified part of the selected field structure remained above the sustainment condition. Amplitude modulation, therefore, controlled both plasma sustainment and the active volume within the frequency-selected mode.

            This result completes the pulse shaping picture. Frequency selects the resonant field structure and breakdown location, peak power initiates breakdown, and reduced incident power can then sustain the selected plasma state with lower average power delivered to the dielectric assembly. Pulse shaping, therefore, provides a route to ignite a localized plasma near the dielectric surfaces and maintain it with reduced resonator heating.

    \subsection{Temporal Evolution of Mode Selected Microwave Breakdown} \label{results:spatiotemporal}
    
        The frequency and amplitude modulation experiments used time-integrated images to identify which plasma structures appeared under different excitation conditions. Those measurements, however, could not determine whether multi-region resonant patterns represented simultaneous microplasmas or self-emission from the plasma moving between different resonant zones within the plasma. High-speed imaging synchronized with microwave power measurements was used to resolve plasma evolution within individual microwave pulses. These experiments addressed two questions. First, they tested whether distinct microplasmas could form simultaneously within a selected resonant field structure or if the resonances would compete with one another as plasmas formed. Second, they tested larger gap conditions, where a reduced quality factor broadened the resonant features and caused neighboring modes to blend more. This determined whether microplasmas remained localized within the selected field structure or redistributed between competing resonant solutions during a single pulse.
        
        Two gap conditions were examined. The first used a moderately increased gap to test whether selected \modebeta and \modegamma structures could still ignite multiple localized plasma regions and sustain them during a single pulse. The second used a larger gap to test whether reduced modal selectivity allowed the plasma to redistribute between regions of a more strongly superimposed field structure. Together, these measurements determine when the mode-selected breakdown remains spatially stable and when plasma loading drives redistribution during a single microwave pulse.

        \subsubsection{Single Pulse Evolution at Reduced Modal Amplification} \label{results:spatiotemporal:alpha_beta}
    
            The first high-speed experiment tested whether selected \modebeta and \modegamma field structures could produce distinct microplasmas at the same moment in time. This test was performed at a moderately increased gap, where the modal quality factor and gap field amplification were expected to be lower than in the smaller gap measurements. The purpose was to determine whether the selected field maps could still ignite multiple localized plasma regions and whether those regions remained sustained after plasma loading began.
            
\begin{figure}[t!]
                \centering
                \includegraphics[width=1\textwidth]{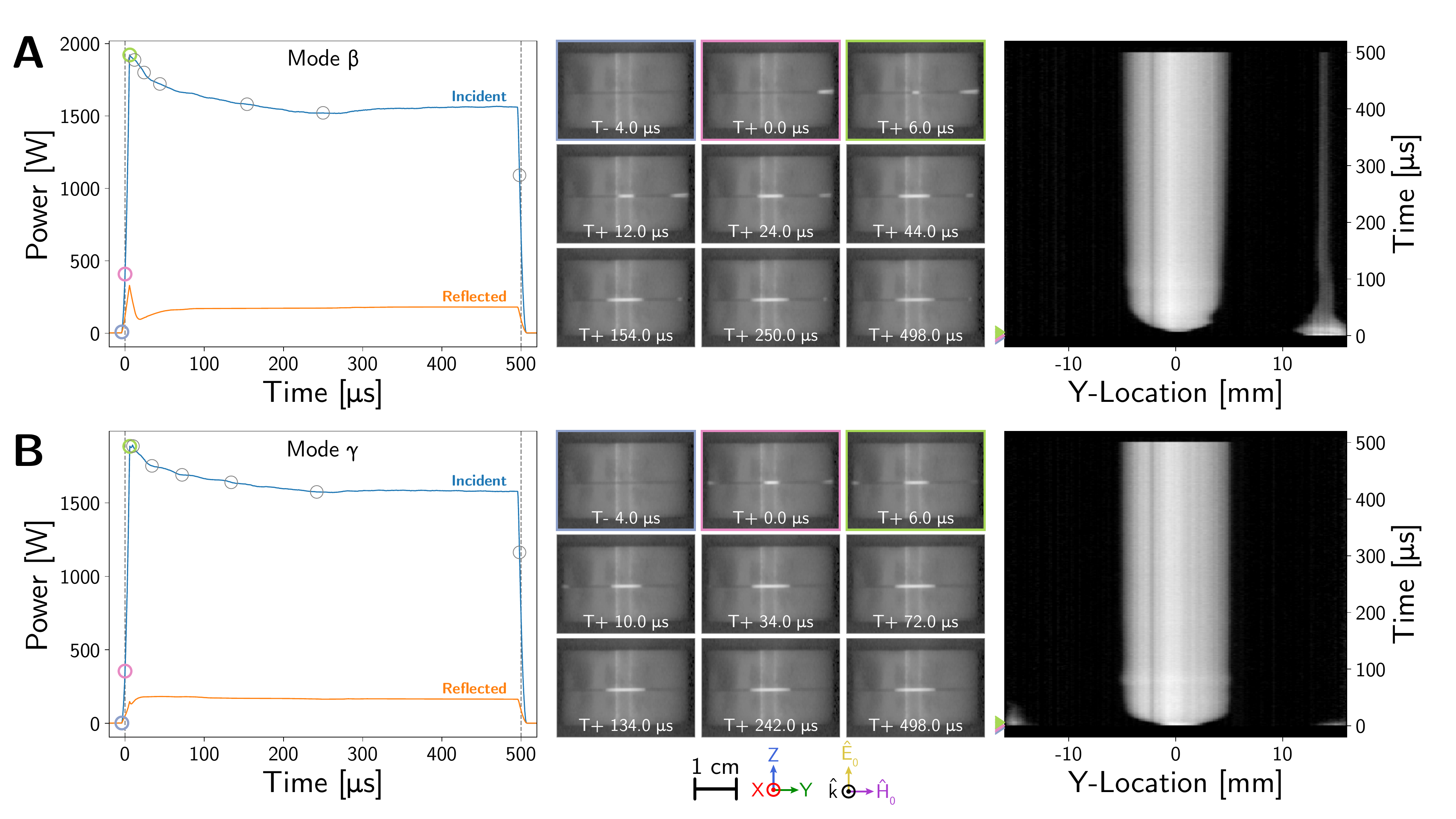}
                \vspace{-25pt}
                \caption{Single pulse evolution of the \modebeta and \modegamma plasma structures measured using synchronized microwave power diagnostics and high speed imaging. The left plots show the incident and reflected powers during the 500~\us microwave pulse. The center images show selected frames acquired near and after plasma ignition. The right panels show streak plots constructed from the 256-frame image sequences at 1800 ns exposure and 500,000 fps, which illustrate the plasma evolution along the inter-resonator gap.}
                \label{ResultsFig6}
            \end{figure}

            The inter-resonator gap was increased to approximately 225~\um, and the resonators were heated to approximately 70~\degC to shift the resonant response into the available excitation range. A 500~\us pulse was used with a 400~Hz PRF and 20\% duty cycle so that the plasma could be tracked after ignition. Incident and reflected microwave powers were recorded synchronously with HPV-X2 images acquired at 500,000~\fps with 1800~ns exposure. The \modebeta and \modegamma conditions were excited at 2.810 and 2.825~GHz, respectively. Selected frames show representative plasma distributions, and streak plots were constructed by extracting the gap region from each frame and stacking the lineouts in time (Fig. \ref{ResultsFig6}).
                    
            The \modebeta condition showed simultaneous plasma formation at more than one location in the gap. Emission appeared near the center and the right side of the inter-resonator gap, and both regions persisted through the pulse. This demonstrates that a selected resonant field structure can support distinct localized microplasmas at the same time, rather than forcing one breakdown site to immediately dominate the others.

            The \modegamma condition also produced multiple emission regions during the initial breakdown period. However, the side regions weakened as the pulse progressed, and the plasma became increasingly concentrated near the center of the gap. This shows that simultaneous ignition does not guarantee equal sustainment. Multiple microplasmas can form within the same selected field map, but weaker regions may fall below the sustainment condition as the plasma-loaded state evolves.
            
            Together, these measurements show that the multi-region plasma patterns observed in time-integrated images were not simply caused by a single plasma moving between locations during the exposure. Distinct microplasmas can form simultaneously within one resonant field structure. The degree to which they remain sustained depends on how strongly each local field region is maintained after plasma loading begins, since field structure can fall below the sustainment condition during the pulse.
            
        \subsubsection{Modal Superposition at Larger Gap Spacing} \label{results:spatiotemporal:trifurcation}
                
            The second high-speed experiment tested the larger gap limit, where the lower quality factor broadened the resonant features and caused neighboring modes to blend more strongly. Under these conditions, the excitation frequency was not expected to isolate a single field structure as cleanly. The experiment, therefore, tested whether microplasmas remained localized within fixed regions of the gap or redistributed between competing resonant solutions during a single pulse.
            
\begin{figure}[H]
                \centering
                \includegraphics[width=1\textwidth]{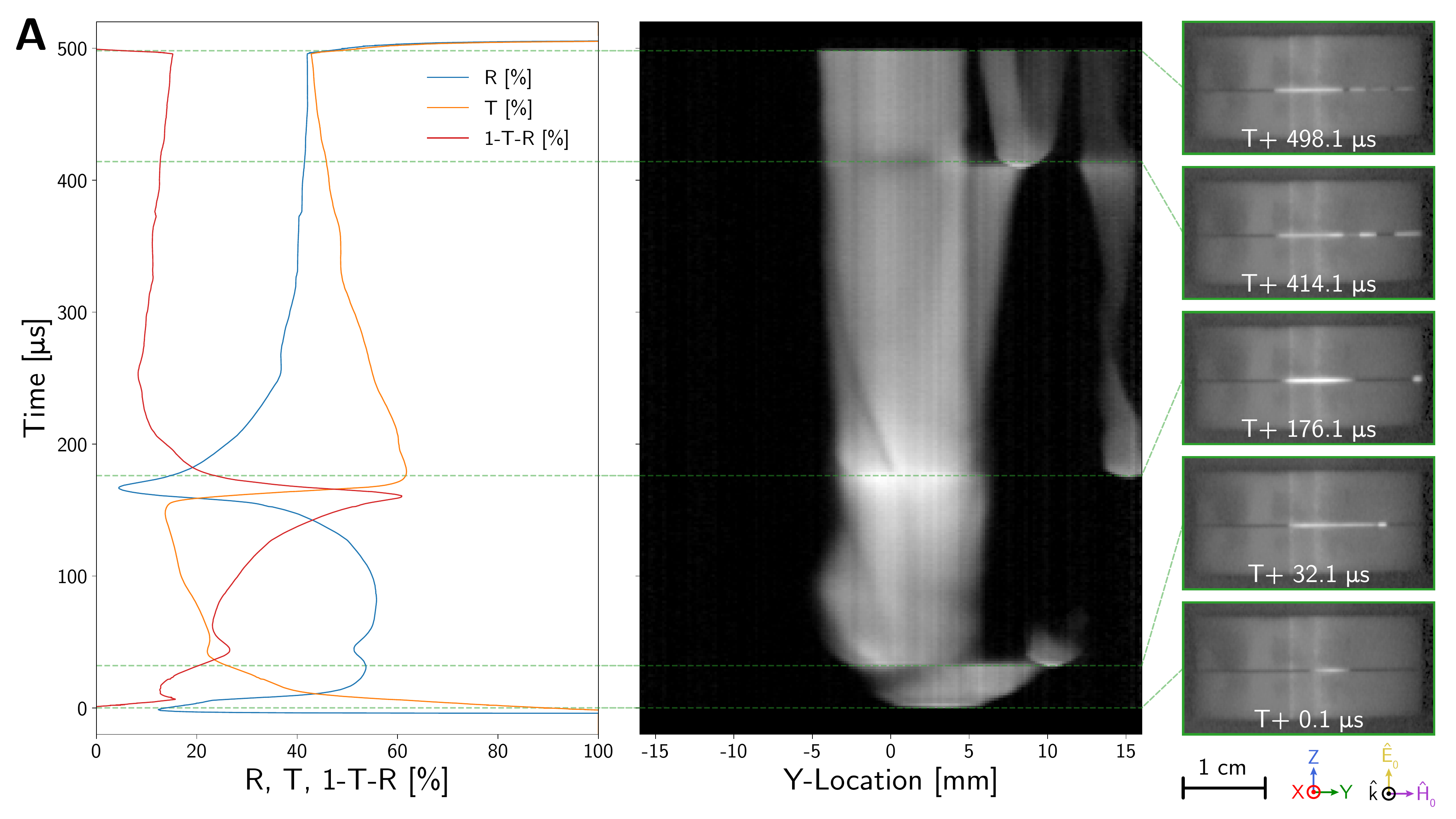}
                \vspace{-25pt}
                \caption{Time-dependent plasma redistribution observed at an excitation frequency of 2.934~GHz, an inter-resonator gap of approximately 300~\um, and a resonator temperature of approximately 80~\degC. The microwave source used a 400~Hz PRF and 20\% duty cycle to create a 500~\us pulse width. The left panel shows synchronized reflected, transmitted, and absorbed fractions of the incident microwave power. The center panel shows a streak plot of the plasma evolution along the inter-resonator gap. The right panel shows selected high-speed images at 1800 ns exposure illustrating changes in plasma location and intensity during the pulse.}
                \label{ResultsFig7}
            \end{figure}

            The inter-resonator gap was increased to approximately 300~\um, the resonator temperature was raised to approximately 80~\degC, and the excitation frequency was fixed at 2.934~GHz. These conditions placed the system near the overlap of the \modebeta and \modegamma responses. High-speed imaging was synchronized with reflected, transmitted, and absorbed microwave power measurements. The camera operated at 500,000~\fps with 1800~ns exposure, and the microwave source used a 400~Hz PRF and 20\% duty cycle to produce a 500~\us pulse. A streak plot was constructed from the gap region to track plasma location and intensity throughout the pulse (Fig. \ref{ResultsFig7}).
                                
            Under this larger gap condition, the plasma did not remain fixed to one region of the gap. Emission shifted between neighboring locations, with one region strengthening as another weakened. The streak plot showed that this redistribution persisted over much of the pulse, indicating that the effect was not only an ignition transient. The simultaneous changes in microwave response show that this motion was coupled to changes in resonator loading.
            
            This behavior indicates that the lower-Q, broader resonant response allowed the \modebeta and \modegamma solutions to compete during the pulse. As the modes became less spectrally distinct, small changes in plasma loading, local heating, alignment, or dielectric temperature could move the system from one favored field distribution to another. In this regime, the plasma did not simply occupy a fixed frequency-selected location. Instead, it redistributed between regions of the blended modal response as the plasma-loaded state evolved.
            
            Together, the high-speed measurements show why pulse shaping must account for temporal stability as well as ignition. At the moderately increased gap, distinct microplasmas formed simultaneously within a selected field structure, demonstrating that multi-region patterns were not caused only by motion during time-integrated exposure. At the larger gap, reduced quality factor broadened the resonant features and caused neighboring modes to blend, allowing the plasma to move between competing solutions during a single pulse. Pulse shaping, therefore, cannot only select a mode but also exceed the breakdown threshold. It must also provide enough modal quality factor, field amplification, and sustainment power to keep the intended plasma regions active after ignition.

\section{Conclusion}

    This work established resonant dielectric field enhancement as a mechanism for spatially controlled microwave breakdown. The dielectric resonators formed standing wave field patterns, while coupling between resonators concentrated the external electric field in the inter-resonator gap until atmospheric pressure breakdown occurred. The results show that resonator geometry defines the available field localization sites, excitation frequency selects among those resonant structures, and the incident pulse shape determines whether the selected structure ignites, how rapidly plasma forms, and whether it remains sustained. Time-resolved imaging further showed that multi-site plasma patterns are not always static. Separate microplasmas can ignite within one resonant field structure, but plasma loading can also redistribute microwave power and cause weaker regions to extinguish or relocate. The plasma-free field, therefore, determines where breakdown can begin, while the plasma-loaded field state determines which structures survive.

    These observations close the loop between resonator design and pulse waveform control, defining a pulse design space for resonant microwave plasma generation. Frequency selects the plasma forming mode, peak power controls access to breakdown and sustainment, pulse timing sets the time available for ignition and equilibration, and average power influences dielectric heating and the longer time discharge state. This framework allows localized microwave breakdown to be designed through the electromagnetic response of the resonator system rather than by electrode placement, fixed applicator geometry, gas flow, or uncontrolled thermal evolution.
    
    The approach also provides a basis for tailoring microwave microplasmas through the design and geometric packing of dielectric materials. The resonant spectrum can be adjusted through dielectric geometry, material properties, coupling gap, and electromagnetic boundary conditions, allowing different plasma locations and structures to be accessed within a fixed hardware configuration. Although this work used a coupled pair of cylindrical resonators, the same design logic extends to other dielectric elements, coupled resonator systems, and structured materials. This capability is relevant to applications where non-equilibrium energy deposition must be localized near a surface, flow, or material interface before excited species relax into bulk gas heating. Resonant dielectric field enhancement could therefore support plasma-assisted chemistry, surface treatment, materials processing, ignition, flow control, and electrothermal propulsion by providing plasma placement set by geometry, structure selected by frequency, and sustainment controlled by power.

\newpage
\ack{This research is supported by the Office of Naval Research grant
N00014-23-1-2306, with Ryan Hoffman as Program Manager.}

\roles{
    \setlength{\parindent}{0pt}
    \setlength{\parskip}{0pt}
    
    \textbf{Arnav Mohapatra} \orcid{0009-0003-2513-6924}\par
        Conceptualization (equal), Data curation (equal), Formal analysis (equal),
        Methodology (equal), Visualization (lead), Investigation (equal),
        Software (lead), Writing--original draft (equal),
        Writing--review \& editing (equal).
    
    \vspace{5mm}
    
    \textbf{Joshua K. Goodrich} \orcid{0009-0009-8520-6027}\par
        Conceptualization (supporting), Data curation (equal), Formal analysis (equal),
        Methodology (supporting), Visualization (supporting), Investigation (equal),
        Software (supporting), Writing--original draft (equal),
        Writing--review \& editing (equal).

    \vspace{5mm}
    
    \textbf{Usman Humayun} 
         Data curation (supporting), Investigation (supporting), Software (supporting).
    
    \vspace{5mm}
    
    \textbf{Thomas C. Underwood} \orcid{0000-0001-5720-2568}\par
        Conceptualization (equal), Formal analysis (supporting),
        Funding acquisition (lead), Investigation (supporting),
        Project administration (lead), Supervision (lead),
        Writing--review \& editing (equal).
}

\printbibliography
\end{document}